\documentclass[12pt]{article}
\usepackage[cp1251]{inputenc}
\usepackage{amsfonts}
\usepackage[english]{babel}
\usepackage{eufrak}
\usepackage{cite}
\usepackage{amssymb}

\begin{document}

\newcommand{\gguide}{{\it Preparing graphics for IOP journals}}
\title{On Non Ergodic Property of Bose Gas with Weak Pair Interaction}

\author{D.V. Prokhorenko \footnote{Institute of Spectroscopy, RAS 142190 Moskow Region, Troitsk, prokhordv@yandex.ru}}
\maketitle

\begin{abstract}
In this paper we prove that Bose gas with weak pair interaction is
non ergodic system. In order to prove this fact we consider the
divergences in some nonequilibrium diagram technique. These
divergences are analogous to the divergences in the kinetic
equations discovered by Cohen and Dorfman. We develop the general
theory of renormalization of such divergences and illustrate it
with some simple examples. The fact that the system is non ergodic
leads to the following consequence: to prove that the system tends
to the thermal equilibrium we should take into account its
behavior on its boundary. In this paper we illustrate this thesis
with the Bogoliubov derivation of the kinetic equations.

\textbf{2000 MSC: 81Q30 (Feynmann integrals and graps)}
\end{abstract}
\newpage
\textit{ \hspace{7cm}Dedicated to the memory of my}

  \textit{\hspace{7.2cm}father V.D. Prokhorenko.}
 \section{Introduction} In this paper we study divergences in some nonequilibrium diagram technique
 which is analogous to the Keldysh diagram technique. It is more or less evident that these divergences are the same
 as the divergences in the kinetic equations discovered by Cohen and Dorfman \cite{1,111,2}. We develop the general theory
 of renormalization of such divergences analogously to the Bogoliubov --- Parasiuk \(R\)-operation method \cite{3,4,5}. Our main
 result can be formulated as follows: for a wide class of Bose systems in
the sense of formal power series on coupling constant there exists
non-Gibbs functional \(\langle \cdot \rangle\), commuting with the number of particle operator such that the
correlators
\begin{eqnarray}
\langle\Psi^\pm(t,{x}_1)...\Psi^\pm(t,{x}_n)\rangle \nonumber
\end{eqnarray}
are translation invariant, do not depend on \(t\) and satisfy
the weak cluster property. Here \(\Psi^\pm\) are the fields
operators and the weak cluster property means the following
\begin{eqnarray}
\lim_{|{a}|\rightarrow\infty}\int \limits_{{R}^{3n}}
\langle\Psi^\pm(t,{x}_1+\delta_1 e_1{a})
...\Psi^\pm(t,{x}_n+\delta_n e_1 {a} )\rangle f({x}_1,...,{x}_n)d^3x_1...d^3x_n\nonumber\\
=\int \limits_{{R}^{3n}} \langle\Psi^\pm(t,{x}_{i_1})...\Psi^\pm(t,{x}_{i_k})\rangle
\langle\Psi^\pm(t,{x}_{i_k})...\Psi^\pm(t,{x}_{i_n})\rangle
\nonumber
\times f({x}_1,...,{x}_n)d^3x_1...d^3x_n,
\end{eqnarray}
there \(\delta_i\in\{1,0\},\;i=1,2...n\) and
\begin{eqnarray}
i_1<i_2<...<i_k,\nonumber\\
i_{k+1}<i_{k+2}<...<i_n,\nonumber\\
\{i_1,i_2,...,i_k\}=\{i=1,2...n|\delta_i=0\}\neq\emptyset,\nonumber\\
\{i_{k+1},i_{k+2},...,i_n\}=\{i=1,2...n|\delta_i=1\}\neq\emptyset.\nonumber
\end{eqnarray}
\(f({x}_1,...,{x}_n)\) is a test function, \(e_1\) is a unit vector parallel to the \(x\)-axis.
 This statement is a simple consequence of the theorem from the section 6.

Let us prove that the existence of such functionals implies
non-ergodic property of the system. We consider the problem only
on classical level. The accurate consideration for the quantum
case can be found in section 10.  Suppose that our system is
ergodic, i.e. there are no first integrals of the system except
energy. Then, the distribution function depends only on energy. We
can represent the distribution function \(f(E)\) as follows:
\begin{eqnarray}
f(E)=\sum c_\alpha \delta(E-E_\alpha),\nonumber
\end{eqnarray}
where the sum can be continuous (integral). Let 1 be some enough
large but finite subsystem of our system. Let 2 be a subsystem
obtained from 1 by translation on the vector \(\vec{l}\) of
sufficiently large length parallel to the \(x\)-axis. Let 12 be a
union of the subsystems 1 and 2. Let \(\rho_1\), \(\rho_2\) and
\(\rho_{12}\) be distribution functions for the subsystems 1, 2
and 12 respectively. Let \(\Gamma_1\), \(\Gamma_2\) and
\(\Gamma_{12}\) be points of the phase spaces for the subsystem 1,
2 and 12 respectively. By the same method as the method used for
the derivation of the Gibbs distribution we find:
\begin{eqnarray}
\rho_{12}=\sum c_\alpha
d_\alpha\frac{e^{-\frac{E_{\Gamma_1}}{T_\alpha}}}{Z_\alpha}\frac{e^{-\frac{E_{\Gamma_2}}{T_\alpha}}}{Z_\alpha},\;d_\alpha>0\,\forall\alpha \nonumber
\end{eqnarray}
 in the obvious notation. But the weak cluster property implies that
 \begin{eqnarray}
 \rho_{12}=\rho_1\rho_2.\nonumber
 \end{eqnarray}
 Therefore all the coefficients \(c_\alpha\) are equal to zero except one. We find that
 \begin{eqnarray}
 f(E)=c\delta(E-E_0)\nonumber
 \end{eqnarray}
 for some constants \(c\) and \(E_0\). So each finite subsystem of our system can be described by Gibbs
 formula and we obtain a contradiction.

Non-ergodic property means that there is no thermalization in infinite Bose-gas system.

This fact implies to prove that the system tends to thermal
equilibrium we should take into account the behavior of the
system on its boundary. Indeed if a system has no boundary the
system is infinite.

To illustrate this fact we will study Bogoliubov derivation of
kinetic equations \cite{6}. When one derives BBGKI-chain one neglects some boundary terms.
If one takes into account this boundary terms
and uses the Bogoliubov method of derivation of the kinetic equations one
finds that these boundary terms compensate the scattering integral.

I think that the dependence of behavior of the system of boundary
can be observed for small systems such as nanosystems or
biological systems.

Note that our main result is closely related with so-called the
Prigogin hypothesis which states that the infinite dimensional
Liouville dynamics can not be derived from the Hamilton dynamics.
The Prigogin hypothesis is proven in \cite{7}

The paper is organized as follows. In section 2 we introduce the
notion of the algebra of canonical commutative relations and
develop an useful representation for some class of the states on
this algebra. In section 3 we describe the von Neumann dynamics
for the states. In section 4 we describe an useful representation
for the von Neumann dynamics --- the dynamics of correlations. In
section 5 we describe the decomposition of the kinetic evolution
operator by so-called trees of correlatios. In section 6 we
describe the general form of the counterterms which subtract the
divergences in the nonequilibrium perturbation theory. In section
7 we describe so-called Friedrichs diagrams. In section 8 we
describe the Bogoliubov---Parasiuk prescriptions and formulate our
main theorem. In section 9 we prove our main theorem. In section
10 we derive the non-ergodic property of Bose gas with weak pair
interaction from our main result. In section 11 we consider one
example related to our general theory. In section 12 we reconsider
the Bogoliubov derivation of the Boltzmann equation. This example
illustrates the main thesis of this paper: to prove that the
system tends to the thermal equilibrium one has to take into
account its behavior on its boundary. Section 13 is a conclusion.

\section{The Algebra of Canonical Commutative Relations}
Let \(S( \mathbb{R}^3)\) be a Schwatrz space of test functions
(infinitely-differentiable functions decaying at infinity faster
than any inverse polynomial with all its derivatives). The algebra
of canonical commutative relations \(\mathcal{C}\) is an unital
algebra generated by symbols \(a^+(f)\) and \(a(f)\) \(f \in S(
\mathbb{R}^3)\) satisfying the following canonical commutative
relations:

a) \(a^+(f)\) is a linear functional of \(f\),

b) \(a(f)\) is an antilinear functional of \(f\),

\begin{eqnarray}
{[a(f),a(g)]}={[a^+(f),a^+(f)]}=0,\nonumber\\
{[a(f),a^+(g)]}=\langle f,g\rangle,\nonumber
\end{eqnarray}
where \( \langle f,g\rangle\) is a standard scalar product in
\(L^2(\mathbb{R}^3)\),
\begin{eqnarray}
\langle f,g\rangle:=\int d^3 x f^*(x)g(x).\nonumber
\end{eqnarray}
Let \(\rho_0\) be a Gauss state on \(\mathcal{C}\) defined by the
following correlator
\begin{eqnarray}
\rho_0(a^+(k)a^+(k'))=\rho_0(a(k)a(k'))=0,\nonumber\\
\rho_0(a^+(k)a(k'))=n(k)\delta(k-k'),\nonumber
\end{eqnarray}
where \(n(k)\) is a real-valued function from the Schwartz space.
In the case then
\begin{eqnarray}
n(k)=\frac{e^{-\beta(\omega(k)-\mu)}}{1-e^{-\beta(\omega(k)-\mu)}},\nonumber
\end{eqnarray}
where \(\mu \in \mathbb{R}\), \(\mu<0\), \(\rho_0\) is called the
Plank state. Here \(\omega(k)=\frac{k^2}{2}\).

Let \(\mathcal{C}'\) be a space of linear functionals on
\(\mathcal{C}\), and \(\mathcal{C}'_{+,1}\) be a set of all states
on \(\mathcal{C}\). Let us make the GNS construction corresponding
to the algebra \(\mathcal{C}\) and the Gauss state \(\rho_0\). We
obtain the set \((\mathcal{H},D,\hat{},\rangle)\) consisting of
the Hilbert space \(\mathcal{H}\), the dense linear subspace \(D\)
in \(\mathcal{H}\), the representation \(\hat{}\) of
\(\mathcal{C}\) by means of the linear operators from \(D\) to \(D\), and the
cyclic vector \(\rangle \in D \), i.e. the vector such that
\(\hat{C}\rangle=D\). This set satisfies the following condition:
\(\forall a \in \mathcal{C}\) \(\langle\hat{a}\rangle=\rho_0(a)\).
Below we will omit the symbol \(\hat{}\), i.e. we will write \(a\)
instead of \(\hat{a}\).

Let us introduce the field operators:
\begin{eqnarray}
\Psi(x)=\frac{1}{(2\pi)^{\frac{3}{2}}} \int
e^{ikx}a(k)dk,\nonumber\\
\Psi^+(x)=\frac{1}{(2\pi)^{\frac{3}{2}}} \int e^{-ikx}a^+(k)dk.\nonumber
\end{eqnarray}

We say that the state \(\rho\) on \(\mathcal{C}\) satisfies the
weak cluster property if

\begin{eqnarray}
\lim_{{a}\rightarrow\infty}\int
\langle\Psi^\pm(t,{x}_1+\delta_1e_1{a})
...\Psi^\pm(t,{x}_n+\delta_n e_1{a})\rangle f({x}_1,...,{x}_n)d^3x_1...d^3x_n\nonumber\\
=\int \langle\Psi^\pm(t,{x}_{i_1})...\Psi^\pm(t,{x}_{i_k})\rangle
\langle\Psi^\pm(t,{x}_{i_k})...\Psi^\pm(t,{x}_{i_n})\rangle
\nonumber\\
\times f({x}_1,...,{x}_n)d^3x_1...d^3x_n,\nonumber
\end{eqnarray}
where \(\delta_i\in\{1,0\},\;i=1,2...n\) and
\begin{eqnarray}
i_1<i_2<...<i_k,\nonumber\\
i_{k+1}<i_{k+2}<...<i_n,\nonumber\\
\{i_1,i_2,...,i_k\}=\{i=1,2...n|\delta_i=0\}\neq\emptyset,\nonumber\\
\{i_{k+1},i_{k+2},...,i_n\}=\{i=1,2...n|\delta_i=1\}\neq\emptyset.\nonumber
\end{eqnarray}
\(f({x}_1,...,{x}_n)\) is a test function. \(e_1\) is an unit vector parallel to the \(x\)-axis.

 \textbf{Definition} The vector of
the form
\begin{eqnarray}
\int v(p_1,...,p_n)a^{\pm}(p_1)...a^{\pm}(p_n)\rangle
d^3p_1...d^3p_n,\nonumber\\
v(p_1,...,p_n) \in S(\mathbb{R}^{3n}). \label{fin}
\end{eqnarray}
is called a finite vector. The finite linear combination of the
vectors of the form (\ref{fin}) is also called a finite vector.

Let \(f(x_1,...,x_k|y_1,...,y_l|v_1,...,v_m|w_1,...,w_n)\) be a
function of the form
\begin{eqnarray}
f(x_1,...,x_k|y_1,...,y_l|v_1,...,v_m|w_1,...,w_n)\nonumber\\
=g(x_1,...,x_k|y_1,...,y_l|v_1,...,v_m|w_1,...,w_n)\nonumber\\
\times\delta(\sum \limits_{i=1}^k x_i-\sum\limits_{j=1}^m w_j-
\sum\limits_{f=1}^l y_f+\sum\limits_{g=1}^n v_g),\nonumber
\end{eqnarray}
where \(g\) is a function from Schwartz space.

Consider the following functional on \(\mathcal{C}\)
\begin{eqnarray}
\rho_f(A):=\int \prod \limits_{i=1}^k dx_i\prod \limits_{j=1}^l
dx_j \prod \limits_{f=1}^m dv_f \prod \limits_{g=1}^n dw_g \nonumber\\
\times f(x_1,...,x_n|y_1,...,y_l|v_1,...,v_m|w_1,...,w_n)\nonumber\\
\times
\rho_0(:a(x_1)...a(x_n)\overbrace{a^+(y_1)...a^+(y_l):A:a(v_1)...a(v_m)}a^+(w_1)...a^+(w_n):).\nonumber
\label{7}
\end{eqnarray}
Here the symbol
\begin{eqnarray}
:(.\overbrace{.):A:(.}.):\nonumber
\end{eqnarray}
means that when one transforms the previous expression to the
normal form according to the Gauss property of \(\rho_0\) one must
neglect all correlators \(\rho_0(a^\pm(x_1)a^\pm(x_n))\) such that
\(a^\pm(x_1)\) and \(a^\pm(x_n)\) both do not come from \(A\).

Let \(\widetilde{\mathcal{C}'}\) be a subspace in \(\mathcal{C'}\)
spanned on the functionals just defined.

Now let us introduce an useful method for the representation of
the states just defined.

Let \(\mathcal{C}_2=\mathcal{C}_+\otimes\mathcal{C}_-\), where
\(\mathcal{C}_+\) and \(\mathcal{C}_-\) are the algebras of
canonical commutative relations. The algebras
\(\mathcal{C}_{\pm}\) are generated by the generators
\(a_\pm(k),a^+_\pm(k)\) respectively satisfying the following
relations:
\begin{eqnarray}
{[a^+_+(k),a^+_+(k')]}={[a_+(k),a_+(k')]}=0,\nonumber\\
{[a^+_-(k),a^+_-(k')]}={[a_-(k),a_-(k')]}=0,\nonumber\\
{[a_+(k),a^+_+(k')]}=\delta(k-k'),\nonumber\\
{[a_-(k),a^+_-(k')]}=\delta(k-k'),\nonumber\\
{[a^{\pm}_+(k),a^{\pm}_-(k)]}=0.\nonumber
\end{eqnarray}
Here we put by definition \(a^-_{\pm}:=a_{\pm}\). Let us consider
the following Gauss functional \(\rho_0'\) on \(\mathcal{C}_2\)
defined by its two-point correlator
\begin{eqnarray}
\rho_0'(a^{\pm}_-(k)a^{\pm}_-(k'))=\rho_0(a^{\pm}(k)a^{\pm}(k')),\nonumber\\
\rho_0'(a^{\pm}_+(k)a^{\pm}_+(k'))=\rho_0(a^{\mp}(k')a^{\mp}(k)),\nonumber\\
\rho_0'(a^+_+(k)a^-_-(k'))=\rho_0'(a^-_+(k)a^+_-(k'))=0, \nonumber\\
\rho_0'(a^-_+(k)a^-_-(k'))=n(k)\delta(k-k'),\nonumber\\
\rho_0'(a^+_+(k)a^+_-(k'))=(1+n(k))\delta(k-k').\nonumber
\end{eqnarray}
One can prove that the functional \(\rho_0'\) is a state.

 Let us make the GNS construction corresponding to the state \(\rho_0'\)
and the algebra \(\mathcal{C}_2\). We obtain the set
\((\mathcal{H}',\tilde{D},\hat{},\rangle)\) consisting of the
Hilbert space \(\mathcal{H}'\), the dense linear subspace
\(\tilde{D}\) in \(\mathcal{H}'\), the representation \(\hat{}\)
of \(\mathcal{C}_2\) by means of the linear operators from
\(\tilde{D}\) to \(\tilde{D}\), and the cyclic vector \(\rangle
\in \tilde{D} \), i.e. the vector such that
\(\hat{C}\rangle=\tilde{D}\). This set satisfies the following
condition: \(\forall a \in \mathcal{C}_2\)
\(\langle\hat{a}\rangle=\rho_0'(a)\). Below we will omit the
symbol \(\hat{}\), i.e. we will write \(a\) instead of
\(\hat{a}\).

Now we can rewrite the functional, defined in (\ref{7})
\(\rho_f\) as follows
\begin{eqnarray}
\rho_f(A)=\langle A'S_f\rangle,\nonumber
\end{eqnarray}
where \(A'\) is an element of \(\mathcal{C}_2\) such that it
contains only the operators \(a_-,a_-^+\) and can be represented
through \(a_-,a_-^+\) in the same way as \(A\) can be represented
through \(a,a^+\). \(S_f\) is an element of \(\mathcal{C}_2\) of the
form
\begin{eqnarray}
S_f=\int \prod \limits_{i=1}^k dx_i\prod \limits_{j=1}^l
dx_j \prod \limits_{f=1}^m dv_f \prod \limits_{g=1}^n dw_g \nonumber\\
\times f(x_1,...,x_n|y_1,...,y_l|v_1,...,v_m|w_1,...,w_n)\nonumber\\
\times
:a^+_+(x_1)...a^+_+(x_n)a_+(y_1)...a_+(y_l)a_-(v_1)...a_-(v_m)a^+_-(w_1)...a^+_-(w_n):.
\label{23}
\end{eqnarray}
Here the symbol \(:...:\) is a normal ordering with respect to the
state \(\rho_0'\).

Denote by \(\tilde{D}'\) the space dual to  \(\tilde{D}\). We just
construct the injection from \(C'\) into \(\tilde{D}'\). Denote
its image by \(\tilde{\mathcal{H}'}\).

By definition the space \(\mathcal{C}''\) is a space of all
functionals on \(\mathcal{C}\) which can be represented as finite
linear combinations of the following functionals
\begin{eqnarray}
\rho(A)=\langle A':S_{f_1}...S_{f_n}:\rangle.\nonumber
\end{eqnarray}
 Here \(A'\) is an element of \(\mathcal{C}_2\) such that it contains only the
 operators \(a_-,a_-^+\) and can be represented through \(a_-,a_-^+\) in the
same way as \(A\) can be represented through \(a,a^+\) and
\(S_{f_i}\) are the elements of the form (\ref{23}). Denote by
\(\tilde{\mathcal{H}}''\) the subspace in \(\tilde{D}'\) spanned
on the vectors \(:S_{f_1}...S_{f_n}:\rangle\) (in obvious sense).

There exists an involution \(\star\) on \(\tilde{\mathcal{H}}'\)
defined by the following formula:
\begin{eqnarray}
\{\int \prod \limits_{i=1}^k dx_i\prod \limits_{j=1}^l
dx_j \prod \limits_{f=1}^m dv_f \prod \limits_{g=1}^n dw_g \nonumber\\
\times f(x_1,...,x_n|y_1,...,y_l|v_1,...,v_m|w_1,...,w_n)\nonumber\\
\times
:a^+_+(x_1)...a^+_+(x_n)a_+(y_1)...a_+(y_l)a_-(v_1)...a_-(v_m)a^+_-(w_1)...a^+_-(w_n):\rangle\}^\star
\nonumber\\
 =\int \prod \limits_{i=1}^k dx_i\prod \limits_{j=1}^l
dx_j \prod \limits_{f=1}^m dv_f \prod \limits_{g=1}^n dw_g \nonumber\\
\times f^*(x_1,...,x_n|y_1,...,y_l|v_1,...,v_m|w_1,...,w_n)\nonumber\\
\times
:a^+_-(x_1)...a^+_-(x_n)a_-(y_1)...a_-(y_l)a_+(v_1)...a_+(v_m)a^+_+(w_1)...a^+_+(w_n):\rangle.\nonumber
\end{eqnarray}

We define the involution \(\star\) on \({\rm Hom \mit}
(\tilde{H}',\tilde{H}')\) by the following equation:
\begin{eqnarray}
(a |f\rangle)^\star=a^\star(|f\rangle)^*,\nonumber
\end{eqnarray}
where \(a \in {\rm Hom \mit} (\tilde{H}',\tilde{H}')\) and
\(|f\rangle \in \tilde{H}'\).

We define also the involution \(\star\) on \(\mathcal{C}^2\) by
the following equation:
\begin{eqnarray}
\{\int \prod \limits_{i=1}^k dx_i\prod \limits_{j=1}^l
dx_j \prod \limits_{f=1}^m dv_f \prod \limits_{g=1}^n dw_g \nonumber\\
\times f(x_1,...,x_n|y_1,...,y_l|v_1,...,v_m|w_1,...,w_n)\nonumber\\
\times
:a^+_+(x_1)...a^+_+(x_n)a_+(y_1)...a_+(y_l)a_-(v_1)...a_-(v_m)a^+_-(w_1)...a^+_-(w_n):\}^\star
\nonumber\\ =\int \prod \limits_{i=1}^k dx_i\prod \limits_{j=1}^l
dx_j \prod \limits_{f=1}^m dv_f \prod \limits_{g=1}^n dw_g \nonumber\\
\times f^*(x_1,...,x_n|y_1,...,y_l|v_1,...,v_m|w_1,...,w_n)\nonumber\\
\times
:a^+_-(x_1)...a^+_-(x_n)a_-(y_1)...a_-(y_l)a_+(v_1)...a_+(v_m)a^+_+(w_1)...a^+_+(w_n):,\nonumber
\end{eqnarray}
where \(f(x_1,...,x_k|y_1,...,y_l|v_1,...,v_m|w_1,...,w_n)\) is a
test function of its arguments. Note that the involution on \({\rm
Hom \mit} (\tilde{H}',\tilde{H}')\) extends the involution on
\(\mathcal{C}^2\). We say that the element \(a \in \mathcal{C}_2\)
is real if \(a^\star=a\). The involution on \(\tilde{H}''\) can be
defined by a similar way.

\section{The von Neumann Dynamics}
Suppose that our system is described by the following Hamiltonian
\begin{eqnarray}
H=H_0+\lambda V,\nonumber
\end{eqnarray}
where
\begin{eqnarray}
H_0=\int d^3k (\omega(k)-\mu)a^+(k)a(k)\; \rm and \mit \nonumber\\
V=\int d^3p_1d^3p_2d^3q_1d^3q_2 v(p_1,p_2|q_1,q_2)\nonumber\\
\times\delta(p_1+p_2-q_1-q_2)a^+(p_1)a^+(p_2)a(q_1)a(q_2).\nonumber
\end{eqnarray}
Here the kernel \(v(p_1,p_2|q_1,q_2)\) belongs to the Schwartz
space of test functions. To point out the fact that \(H\) is
represented through the operators \(a^+,\;a^-\) we will write \(
H(a^+,a^-)\).

The von Neumann dynamics takes place in the space
\(\tilde{\mathcal{H}}''\) and is defined by the following
differential equation:
\begin{eqnarray}
\frac{d}{dt}|f\rangle=\mathcal{L}|f\rangle,\nonumber
\end{eqnarray}
where the von Neumann operator has the form
\begin{eqnarray}
\mathcal{L}=-iH(a^+_-,a^-_-)+iH^\dagger(a^+_+,a^-_+),\nonumber
\end{eqnarray}
where we put by definition:
\begin{eqnarray}
(\int \prod \limits_{i=1}^n d p_i \prod \limits_{j=1}^m d q_j
v(p_1,...,p_n|q_1,...,q_m):a^+(p_1)...a^+(p_n)a(q_1)...a(q_n):)^\dagger\nonumber\\
=\int \prod \limits_{i=1}^n d p_i \prod \limits_{j=1}^m d q_j
v(p_1,...,p_n|q_1,...,q_m)^\ast:a^+(p_1)...a^+(p_n)a(q_1)...a(q_n):.\nonumber
\end{eqnarray}
Let us divide the von Neumann operator into the free operator
\(\mathcal{L}\) and the interaction \(\mathcal{L}_{int}\),
\(\mathcal{L}=\mathcal{L}_0+\lambda \mathcal{L}_{int}\), where
\begin{eqnarray}
\mathcal{L}_0=-iH_0(a^+_-,a^-_-)+iH_0^\dagger(a^+_+,a^-_+),\nonumber\\
\mathcal{L}_{int}=-iH_{int}(a^+_-,a^-_-)+iH_{int}^\dagger(a^+_+,a^-_+).\nonumber
\end{eqnarray}
Note that the operators \(\mathcal{L}_0\) and \(\mathcal{L}_1\)
are real (with respect the involution \(\star\)).

 Let us introduce kinetic evolution operator (in the
interaction representation)
\begin{eqnarray}
U(t'',t')=e^{-\mathcal{L}_0t''}e^{\mathcal{L}(t''-t')}e^{\mathcal{L}_0t'}.\nonumber
\end{eqnarray}
After differentiating with respect to \(t\) we find the differential equation
for \(U(t,t')\).
\begin{eqnarray}
\frac{d}{dt}U(t,t')=\mathcal{L}_{int}(t)U(t,t'),\nonumber
\end{eqnarray}
where
\begin{eqnarray}
\mathcal{L}_{int}(t)=e^{-\mathcal{L}_0t}\mathcal{L}_{int}e^{\mathcal{L}_0t}.\nonumber
\end{eqnarray}
So the state \(\rangle_{\rho}\) under consideration in the space \(\tilde{\mathcal{H}}'' \) in the
interaction representation
has the form
\begin{eqnarray}
\rangle_\rho=T \rm exp \mit (\int \limits_{-\infty}^0
\mathcal{L}_{int}(t)dt)\rangle,\nonumber
\end{eqnarray}
where \(T\) is the time-ordering operator.

Note that we have a linear map from \(\tilde{H}''\) into
\(\widetilde{\mathcal{C}}'\). It is easy to see that the von
Neumann dynamics is in agreement with the Heizenberg dynamics in
\(\mathcal{C}'\).

\section{Dynamics of Correlations}
Let us construct some new representation of the von Neumann dynamics useful for
the renormalization program. This representation is called the
dynamics of correlations. The ideas of the dynamics of correlations belongs to I. Prigogin \cite{8}.
 The dynamics of correlations takes place
in the space
\begin{eqnarray}
\mathcal{H}_c:={\bigoplus \limits_0^{\infty} \rm sym \mit
{\otimes}^n} \tilde{\mathcal{H}}'.\nonumber
\end{eqnarray}
Now let us describe how the operators \(\mathcal{L}_0^c\) and
\(\mathcal{L}_{int}^c\) act in the space \(\mathcal{H}_c\).

Let us define the actions of operators \(\mathcal{L}_0^c\) and
\(\mathcal{L}_{int}^c\) which are corresponds to the operators
\(\mathcal{L}_0\) and \(\mathcal{L}_{int}\).

By definition all the spaces \(\otimes^n \tilde{{\mathcal{H}}}'\)
are invariant under the actions of operators \(\mathcal{L}_{0}^c\).
Note that the space \(\tilde{{\mathcal{H}}}'\) is invariant under
the action of operator \(\mathcal{L}_{0}\). Let us denote the
restriction of \(\mathcal{L}_{0}\) to the space
\(\tilde{{\mathcal{H}}}'\) by the symbol \(\mathcal{L}_{0}'\). By
definition the restriction  of \(\mathcal{L}_{0}^c\) to the each
subspace \({\rm sym \mit}\otimes^n \tilde{{\mathcal{H}}}'\) of
\(\mathcal{H}_c\) has the form
\begin{eqnarray}
\mathcal{L}'_0\otimes\mathbf{1}\otimes...\otimes\mathbf{1}+
\mathbf{1}\otimes\mathcal{L}'_0\otimes...\otimes\mathbf{1}+...+
\mathbf{1}\otimes\mathbf{1}\otimes...\otimes\mathcal{L}'_0.\nonumber
\end{eqnarray}
Now let us define \(\mathcal{L}^c_{int}\). Let \(|f\rangle \in
\mathcal{H}_c\), belongs to the subspace \( \otimes^n
\tilde{{\mathcal{H}}}'\) and has the form:
\begin{eqnarray}
|f\rangle=\sum \limits_{i=0}^m f_1^i\rangle\otimes...\otimes
f_n^i\rangle,\nonumber
\end{eqnarray}
where \(f_j^i\) has the form
\begin{eqnarray}
f_i^j=\int \prod \limits_{i=1}^k dx_i\prod \limits_{j=1}^l
dx_j \prod \limits_{f=1}^m dv_f \prod \limits_{g=1}^n dw_g \nonumber\\
\times f(x_1,...,x_n|y_1,...,y_l|v_1,...,v_m|w_1,...,w_n)\nonumber\\
\times
:a^+_+(x_1)...a^+_+(x_n)a_+(y_1)...a_+(y_l)a_-(v_1)...a_-(v_m)a^+_-(w_1)...a^+_-(w_n):.\label{35}
\end{eqnarray}

By definition,
\begin{eqnarray}
\mathcal{L}_{int}^{c,l}|f\rangle=0\nonumber
\end{eqnarray}
if \(l> n\). Let us consider the following vector in
\(\tilde{\mathcal{H}}''\)
\begin{eqnarray}
\sum \limits_{i=1}^{m}:\prod \limits_{j=1}^n f^i_j:\rangle.\nonumber
\end{eqnarray}
Let us transform the expression \(\mathcal{L}_{int} \sum
\limits_{i=1}^m :\prod \limits_{j=1}^n f^i_j:\) to the normal
form. Let us denote by \(h_l\) the sum of all the terms in the
previous expression such that exactly \(l\) operators \(f^i_j\)
couple with \(\mathcal{L}_{int}\).
 We find that \(h_l\rangle\) has the following form
\begin{eqnarray}
h_l\rangle=\sum \limits_{i=1}^k :g_1^i...g_{n-l+1}^i:\rangle\nonumber
\end{eqnarray}
for some \(k\). Here \(g^i_k\) has the form of right hand side of (\ref{35}).
Now let us consider the following vector
\begin{eqnarray}
|f\rangle_l^c=\rm sym \mit \sum \limits_{i=1}^k
:g_1^i:\rangle\otimes...\otimes :g^i_{n-l+1}:\rangle,\nonumber
\end{eqnarray}
where we define symmetrization operator as follows
\begin{eqnarray}
\rm sym \mit (f_1\otimes...\otimes f_n)\nonumber\\
=\frac{1}{n!}\sum \limits_{\sigma \in S_n}
f_{\sigma_1}\otimes...\otimes f_{\sigma(n)}.\nonumber
\end{eqnarray}
(\(S_n\) --- the group of permutation of \(n\) elements.) Put by
definition
\begin{eqnarray}
\mathcal{L}_{int}^{c,l}|f\rangle=|f\rangle_l^c.\nonumber
\end{eqnarray}
One can prove that this definition is correct. Analogously, in the
following expression
\begin{eqnarray}
\mathcal{L}_{int}\sum \limits_{i=1}^m :\prod \limits_{j=1}^n
f^i_j:\rangle\nonumber
\end{eqnarray}
let us keep only the terms such that \(\mathcal{L}_{int}\) does not
couple with any of \(f^i_j\). Let us write the sum of such terms
as follows
\begin{eqnarray}
\sum \limits_{i=1}^f :\prod \limits_{j=1}^{n+1} h^i_j:\rangle.\nonumber
\end{eqnarray}
Here \(h^i_j\rangle\) has the form of right hand side of (\ref{35}).
Let \(|h\rangle\) be a vector in \( {\rm sym \mit \otimes^{n+1}}
\tilde{\mathcal{H}}'\) defined as follows
\begin{eqnarray}
|h\rangle=\rm sym \mit \sum \limits_{i=1}^f
\bigotimes\limits_{j=1}^{n+1}:h^i_j:\rangle.\nonumber
\end{eqnarray}
Put by definition
\begin{eqnarray}
\mathcal{L}_{int}^{c,0}|f\rangle=  |h\rangle.\nonumber
\end{eqnarray}

We have the evident linear map
\(F:\mathcal{H}_c\rightarrow\tilde{\mathcal{H}}''\) which assigns
to each vector \(\rm sym \mit
:f_1:\rangle\otimes...\otimes:f_n:\rangle\) the vector
\(:f_1...f_n:\rangle\). Denote by \(U^c\) the evolution operator
in interaction representation in the dynamics of correlation. The following statement describes the relation between
the von Neumann dynamics and the dynamics of correlations.

\textbf{Statement.} The
following relation holds:
\begin{eqnarray}
F\circ U^c(t',t'')=U(t',t'')\circ F.\nonumber
\end{eqnarray}
\section{The tree of correlations}
The useful representation of dynamics in \(\mathcal{H}_c\) is a
decomposition by so called trees of correlations.

\textbf{Definition.} A graph is a triple \(T=(V,R,f)\), where
\(V\), \(R\) are finite sets called the set of vertices and set of lines
respectively and \(f\) is a map:
\begin{eqnarray}
h:R\rightarrow V^{(2)}\cup V\times \{+\}\cup V\times\{-\},\nonumber
\end{eqnarray}
where \(V^{(2)}\) is a set of all disordered pairs \((v_1,v_2)\),
\(v_1, v_2 \in V\) such that \(v_1\neq v_2\).

If \((v_1,v_2)=f(r)\) for some \(r \in R\) we say that the
vertices \(v_1\) and \(v_2\) are connected by a line \(r\). If
\(f(r)=(v_1,v_2)\), \(v_1,v_2 \in V\) we say that the line \(r\)
is internal.

\textbf{Remark.} We use this unusual definition of graphs only in purpose
of this section to simplify our notations.

\textbf{Definition.} The graph \(\Gamma\) is called connected graph if
for two any vertices  \(v,v'\) there exists a sequence of vertices
\(v=v_0,v_1,...,v_n=v'\) such that
 \(\forall\, i=0,...,n-1\) the vertices
 \(v_i\) and \(v_{i+1}\) are connected by some line.

By definition we say that the line \(r\) is an internal line if
\(f(r)=(v_1,v_2)\) for some vertices \(v_1\) and \(v_2\).

For each graph \(\Gamma\) we define its connected components by
the obvious way.

\textbf{Definition.}  We say that the graph \(\Gamma\) is a tree
or an acyclic graph if the number of its connected components
increases after removing an arbitrary line.

\textbf{Definition.} The elements of the set
\(f^{-1}(V\times\{-\})\) we call the shoots. Put by definition
\(R_{sh}=f^{-1}(V\times\{-\})\). The elements of the set
\(f^{-1}(V\times\{+\})\) we call the roots. Put by definition
\(R_{root}=f^{-1}(V\times\{+\})\).

\textbf{Definition.} Directed tree is a triple \((T,
\Phi_v,\Phi_{sh})\), where \(T\) is a tree and \(\Phi_v\) and
\(\Phi_{sh}\) are the following maps:
\begin{eqnarray}
\Phi_v: V\rightarrow \{1,2,...\sharp V\},\nonumber\\
 \Phi_{sh}: R_{sh}\rightarrow\{1,2,...,\sharp R_{sh}\}.\nonumber
\end{eqnarray}

\textbf{Definition.} We will consider the following two directed
trees \((T,\Phi_v,\Phi_{sh})\) and \((T',\Phi'_v,\Phi'_{sh})\) as
identical if we can identify the sets of lines \(R\) and \(R'\) of
\(T\) and \(T'\) respectively and identify the sets of vertices
\(V\) and \(V'\) of \(T\) and \(T'\) respectively such that after
these identification the trees \(T\) and \(T'\) become the same,
the functions \(\Phi_v\) and \(\Phi'_v\) become the same and the
functions \(\Phi_{sh}\) and \(\Phi'_{sh}\) become the same.

Denote by \(r(T)\) the number of roots of \(T\) and by \(s(T)\)
the number of shoots of \(T\). Below, we will denote each directed
tree \((T,\Phi_v,\Phi_{sh})\) by the same symbol \(T\) as a tree
omitting the reference to \(\Phi_v\), \(\Phi_{sh}\) and write
simply tree instead of the directed tree.

We say that the connected directed tree \(T\) is right if there
exists exactly one line from \(f^{-1}(V\times\{+\})\).

We say that the tree \(T\) is right if each its connected
component is right.

The vertex \(v\) of the tree \(T\) is called a root vertex if
\((v,+) \in f^{-1}(R)\).

To point out the fact that some object \(A\) corresponds to a tree
\(T\) we will often write \(A_T\). For example we will write
\(T=(V_T,R_T,f_T)\) instead of \(T=(V,R,f)\).

\textbf{Definition.} For each connected right tree \(T\) there
exists an essential partial ordering on the set of its vertices.
Let us describe it by induction on the number of its vertices.
Suppose that we have defined this relation for all right trees
such that the number of their vertices is less or equal than
\(n-1\). Let \(T\) be a right tree such that the number of its
vertices is equal to \(n\). Let \(v_{max}\) be a root vertex of
\(T\). Put by definition that the vertex \(v_{max}\) is a maximal
vertex. Let \(v_1,...,v_k\) be all of its children i.e. the
vertices connected with \(v_{max}\) by lines. By definition each
vertex \(v_i<v_{max},i=1,...,k\). We can consider the vertices
\(v_1,...,v_k\) as a root vertices of some directed trees
\(T_i,i=1,...,k\). By definition the set of vertices of \(T_i\)
consists of all vertices \(v\) which can be connected with \(v_i\)
by some path \(v=v'_1,....,v'_l=v_i\) such that \(v_{max}\neq
v'_j\) for all \(j=1,...,l\). The incident relations on \(T_i\)
are induced by incident relations on \(T\). Put by definition that
\(\forall (i,j),\;i,j=1,...,k,\;i\neq j\) and for any two vertices
\(v'_1 \in T_i\) and \(v'_2 \in T_j\) \(v'_1\nless v'_2\). If
\(v_1',v_2' \in T_i\) for some \(T_i\) we put \(v_1'\lessgtr v_2'
\) in \(T\) if and only if \(v_1'\lessgtr v_2'\) in the sense of
ordering on \(T_i\). We put also \(v<v_{max}\) for every vertex
\(v\neq v_{max}\). These relations are enough to define the
partial ordering on \(T\).

If the tree \(T\) has several connected components we define a
partial ordering at each its connected components as previously
and put \(v_1\ngtr v_2\) if \(v_1\) and \(v_2\) do not belongs to
the same connected component of \(T\).

Below without loss of generality we suppose that for each tree
of correlation \(T\) and its line \(r\) the pair
\((v_1,v_2)=f(r)\) satisfies to the inequality \(v_1>v_2\).

\textbf{Definition.} The tree of correlations \(C\) is a triple
\(C=(T,\varphi,\vec{\tau})\), where \(T\) is a directed tree,
\(\vec{\tau}\) is a map from \(R\setminus R_{sh}\) to
\(\mathbb{R}^+:=\{x \in \mathbb{R}|x\geq 0\}\):
\begin{eqnarray}
\vec{\tau}:R\setminus R_{sh}\rightarrow \mathbb{R}^+,\nonumber\\
r\mapsto \tau(r),\nonumber\\
(\tau(r))_{r \in \mathbb{R}}=\vec{\tau}(r),\nonumber
\end{eqnarray}
and \(\varphi\) is a map which assigns to each vertex \(v\) of
\(T\) an element
\begin{eqnarray}
\varphi(v) \in {\rm Hom \mit} ({\bigotimes \limits_{r\rightarrow
v}} {\tilde{\mathcal{H}}', \tilde{\mathcal{H}}'})\nonumber
\end{eqnarray}
of a space of linear maps from \(\bigotimes \limits_{r \rightarrow
v} \tilde{\mathcal{H}}'\) to \(\tilde{\mathcal{H}}'\).

In \(\bigotimes \limits_{(r\rightarrow v)}\tilde{\mathcal{H}}'\)
the tensor product is taken over all lines \(r\) such that
\(r\rightarrow v\). Let \(v\) be a vertex of the tree \(T\). If
\(f(r)=(v',v)\) for some vertex \(v'\) or \(f(r)=(v,+)\) we say
that the line comes from the vertex \(v\) and write \(r\leftarrow
v\). If \(f(r)=(v,v')\) for some vertex \(v'\) or \(f(r)=(v,-)\)
we say that the line comes into the vertex \(v\) and write
\(r\rightarrow v\).

\textbf{Definition.} Let \((T,\varphi,\vec{\tau})\) be a tree of
correlations such that for each vertex \(v\)
\(\varphi(v)=\mathcal{L}_{int}^{c,l_v}\), where \(l_v\) is a
number of lines coming into \(v\). We call this tree the von
Neumann tree and denote it by \(T_{\vec{\tau}}\). We also say that
\(\varphi\) is a von Neumann vertex function.

\textbf{Definition.} To each tree of correlations
\((T,\varphi,\vec{\tau})\) we assign an element
\begin{eqnarray}
U^t_{T,\varphi}(\vec{\tau}) {\in \rm Hom \mit}(\bigoplus
\limits_{R_{sh}} \tilde{\mathcal{H}}',\bigoplus \limits_{R_{root}}
\tilde{\mathcal{H}}')\nonumber
\end{eqnarray}
by the following way:

If \(T\) is disconnected then
\begin{eqnarray}
U^t_{T,\varphi}(\vec{\tau}) f_1\otimes...\otimes f_n \nonumber\\
=\bigotimes \limits_{CT}\{
U^t_{CT,C\varphi}(C\vec{\tau})\bigotimes \limits_{i \in R_{sh}(CT)}
f_i\}.\nonumber
\end{eqnarray}
Here the number of connected components of \(T\) is equal to \(n\),
and connected components of \(T\) are denoted by \(CT\).
\(C\varphi\) and \(C\vec{\tau}\) are the restrictions of
\(\varphi\) and \(\vec{\tau}\) to the sets of vertices and lines
of \(CT\) respectively. \(R_{sh}(CT)\) is a set of shoots of \(CT\).
Now let \(T\) be a connected tree. To define
\begin{eqnarray}
U^t_{T,\varphi}(\vec{\tau})\bigotimes \limits_{r \in R_{sh}} f_r\nonumber
\end{eqnarray}
by induction it is enough to consider the following two cases.

 case 1). The tree \(T\) has no shoots.

 a) Suppose that the tree \(T\) has more than one vertex. Let
 \(v_{min}\) be some minimal vertex of \(T\) and \(v_0\) be a
 vertex such that an unique line \(r_0\) comes from \(v_{min}\)
 into \(v_0\). Let \(T'\) be a tree obtained from \(T\) by
 removing the vertex \(v_{min}\) of \(T\). Let
 \(\vec{\tau}'\) be a restriction of \(\vec{\tau}\) to
 \(R\setminus\{r_0\}\). Let \(\varphi'\) be a function, defined on
 \(V\setminus\{v_{min}\}\) as follows: \(\varphi'(v)=\varphi(v)\)
 if \(v\neq v_0\) and
 \begin{eqnarray}
 \varphi'(v_0)\bigotimes \limits_{r\rightarrow v_0;\;
 r\neq r_0} f_r=\varphi(v_0)\bigotimes \limits_{r\rightarrow v_0} h_r,\nonumber
 \end{eqnarray}
 where
 \begin{eqnarray}
 h_r=f_r\; \rm if \mit \; r\neq r_0,\; \rm and \mit \nonumber\\
 h_{r_0}=e^{\mathcal{L}_0\tau(r_0)} \varphi(v_{min}).\nonumber
 \end{eqnarray}

 Put by definition
 \begin{eqnarray}
 U^t_{T,\varphi}(\vec{\tau})\rangle=U^t_{T',\varphi'}(\vec{\tau}')\rangle,\;\nonumber
\end{eqnarray}

b) The tree \(T\) has only one vertex \(v_{min}\). Then
\begin{eqnarray}
U^t_{T,\varphi}(\vec{\tau})=e^{-(t-\tau)\mathcal{L}_0}\varphi(v_{min}).\nonumber
\end{eqnarray}

Case 2.) The tree \(T\) has a shoot \(r_0\) coming into the vertex
\(v_0\). In this case instead of the tree \((T,\varphi,\vec{\tau})\) we
consider the tree \((T',\varphi',\vec{\tau}')\), where the tree
\(T'\) has the same vertices as \(T\), the set of lines of \(T\)
is obtained by removing the line \(r_0\) from the set of lines of
\(T'\), the function \(\vec{\tau}'\) is a restriction of the
function \(\vec{\tau}\) to the set of lines of \(T'\) and the
function \(\varphi'\) is defined as follows:
\begin{eqnarray}
\varphi'(v)=\varphi(v), \rm if \mit v\neq v_0 \;, \nonumber\\
\varphi'(v_0)\bigotimes \limits_{r\rightarrow v_0\;r\neq r_0} h_r=
\varphi(v_0)\bigotimes \limits_{r\rightarrow v_0}
g_r,\;\rm where \mit \nonumber\\
g_r=h_r,\; \rm if \mit \;r\neq r_0,\; \rm and \mit \nonumber\\
g_r=e^{\mathcal{L}_0(t-t_r)} f_{r_0}.\nonumber
\end{eqnarray}
Here we put
\begin{eqnarray}
t_r=\sum \tau_{r'},\nonumber
\end{eqnarray}
where the sum is taken over all lines \(r'\) which forms decreasing way
coming from \(+\) to \(v_0\).
Put by definition
\begin{eqnarray}
U^t_{T,\varphi}(\vec{\tau})|f\rangle:=U^t_{T',\varphi'}(\vec{\tau'})|f'\rangle,\nonumber
\end{eqnarray}
where
\begin{eqnarray}
|f'\rangle=\bigotimes \limits_{r \in (R_{sh})_{T'}} f_r.\nonumber
\end{eqnarray}

Let \((T,\varphi,\vec{\tau})\) be some tree of correlations. We
can identify the tensor product
\begin{eqnarray}
\bigotimes \limits_{r \in R_{sh}}\tilde{\mathcal{H}}'_r\nonumber
\end{eqnarray}
with
\begin{eqnarray}
\bigotimes \limits_{i=1}^{sh(T)}\tilde{\mathcal{H}}'\nonumber
\end{eqnarray}
and the tensor product
\begin{eqnarray}
\bigotimes \limits_{r \in R_{root}}\tilde{\mathcal{H}}'_r\nonumber
\end{eqnarray}
with
\begin{eqnarray}
\bigotimes \limits_{i=1}^{r(T)}\tilde{\mathcal{H}}.'\nonumber
\end{eqnarray}
Using these identifications let us consider an operator
\(V^t_{T,\varphi}(\vec{\tau}):\mathcal{H}^c\rightarrow
\mathcal{H}^c\) defined by the following formula
\begin{eqnarray}
V^t_{T,\varphi}=\rm sym \mit \circ U^t_{T,\varphi} \circ
P_{sh(T)},\nonumber
\end{eqnarray}
where \(P_{sh(T)}\) is a projection of \(\mathcal{H}_c\) to \(\rm
sym \mit \bigotimes \limits_{i=1}^{sh(T)} \tilde{\mathcal{H}}'\).

\textbf{Remark.} If \((T,\varphi,\vec{\tau})\) is a von Neumann
tree of correlations then we will shortly denote the operators
\(U^t_{(T,\varphi)}\) and \(V^t_{(T,\varphi)}\) by \(U^t_T\) and
\(V^t_T\) respectively.

The following theorem holds:

\textbf{Theorem.} The following representation for the evolution
operators holds (in the sense of formal power series on
coupling constant \(\lambda\)).
\begin{eqnarray}
U^c(t',t'')=\sum \limits_T\frac{\lambda^{n_T}}{n_T!} \int
\limits_{\forall r \in R_{sh}\;t-t_r>t''} V^t_T(\vec{\tau})d
\vec{\tau}.\nonumber
\end{eqnarray}
Here \(n_T\) is a number of vertices of the directed tree \(T\).
\section{The general theory of renormalization of
\(U(t,-\infty)\rangle\)}

In the present section we by using the decomposition of
correlations dynamics by trees describe the general structure of
counterterms of \(U(t,-\infty)\rangle\), which subtract the
divergences from \(U(t,-\infty)\rangle\). We will prove in the section 10 below that there exist
divergences in the theory. Note that the structure of \(R\)-operation for the processes at large times
for some class of systems has been considered at \cite{9}

Let \(T\) be a tree. Let us give a definition of its right
subtree.

\textbf{Definition.} Let \(v_1,...,v_n\) be vertices of \(T\) such
that \(\forall i,j=1,...,n,\;i\neq j\) \(v_i\nless v_j\). Let us
define subtree \(T_{v_1,...,v_n}\). By definition the set of
vertices \(V_{T_{v_1,...,v_n}}\) of \(T_{v_1,...,v_n}\) consists
of all vertices \(v\) such that \(v< v_i\) for some \(i=1,...,n\).

The set \(R_{T_{v_1,...,v_n}}\) of all lines of the tree
\(T_{v_1,...,v_n}\) consists of all lines \(r\) of \(R_T\) such that \(h(r)=(v'',v')\) and \(v',v''\leqslant v_i\) for
some \(i=1,...,n\). The incident relations on \(T_{v_1,...,v_n}\)
are induced by the incident relations of \(T\) except the
following point: if the line \(r\) comes from the vertex \(v\)
into \(v_i,\;i=1,...,n\) we put
\(f_{T_{\{v_1,...,v_n\}}}(r)=(v,+)\). In this case the line \(r\)
is a root and the vertex \(v\) is a root vertex of the tree
\(V_{T_{\{v_1,...,v_n\}}}\). The tree \(T_{v_1,...,v_n}\) is
called a right subtree of \(T\).

\textbf{The Bogoliubov --- Parasiuk  renormalization
prescription.} Let us define the following operator:
\begin{eqnarray}
W_{r_0}(t)=\bigotimes \limits_{r \in R_{root}(T)}Z_{r,r_0}(t),\nonumber
\end{eqnarray}
where by definition,
\begin{eqnarray}
Z_{r,r_0}(t)=1,\;\rm if \mit\; r\neq r_0,\;\rm and \mit \nonumber\\
 Z_{r,r_0}(t)=e^{-\mathcal{L}_0t}.
 \end{eqnarray}
 We say that the amplitudes \(\{A_{T,\varphi}\}\) are time ---
 translation invariant amplitudes if for each tree \(T\) and for
 each its root line \(r_0\)
 \begin{eqnarray}
 W_{r_0}(t)A_{T,\varphi}=A_{T,\varphi}.\nonumber
 \end{eqnarray}

 For each set of amplitudes \(A_{T,\varphi}\) put by definition:
\begin{eqnarray}
A_{T,\varphi}\rangle=F\circ A_{T,\varphi},\nonumber
\end{eqnarray}
where \(T\) is an arbitrary tree without shoots.

Now let us formulate our main result.

\textbf{Theorem.} There exists a procedure called renormalization which to each tree \(T\) without shoots
 assign the amplitudes
 \(\Lambda_{T,\varphi}\) satisfying to the following properties a)-e):

 a) If the tree \(T\) is not connected and \(\{CT\}\) is a set of
 its
 connected components, while \(\{C\varphi\}\) is a set of its
 restriction of \(\varphi\) to \(CT\)
 \begin{eqnarray}
 \Lambda_{T,\varphi}=\bigotimes \limits \Lambda_{CT,C\varphi}\nonumber
 \end{eqnarray}
 in obvious notations.

  b) The amplitudes \(\Lambda_{T,\varphi}\) are real i.e.
 \begin{eqnarray}
 (\Lambda_{T,\varphi})^\star=\Lambda_{T,\varphi^*}\nonumber
 \end{eqnarray}

 c) The amplitude \(\Lambda_{T,\varphi}\)  satisfies the property of
 time-translation invariance.

 It has been proven that
 \begin{eqnarray}
 U(t,-\infty)\rangle=\sum \limits_T \frac{\lambda^{n_T}}{n_T} \int d\vec{\tau}
 U_T^t(\vec{\tau})\rangle. \label{63}
 \end{eqnarray}

In the last formula the summation is taken over all trees \(T\) without shoots.

 Let \(T\) be a tree without shoots and \(T'\) be a right subtree
 of \(T\) in the described before sense. Let us define the
 amplitude
 \begin{eqnarray}
 \Lambda_{T',\varphi}\star U^t_{T,\varphi}(\vec{\tau}).\nonumber
 \end{eqnarray}
Let by definition \(T\backslash T'\) be a tree obtained by
removing from the set \(V_T\) all the vertices of \(T'\) and from
the set \(R_T\) all the internal lines of \(T'\). In (\ref{63})
\(\vec{\tau}\) is a map from \(R_{T\setminus T'}\) into
\(\mathbb{R}^+\).

We can consider the amplitude \(U^t_{T\setminus T'}\) as a map
\begin{eqnarray}
{\bigotimes \limits_{(R_{T\setminus
T'})_{sh}}}\tilde{H}'\rightarrow {\bigotimes
\limits_{(R_{T\setminus T'})_{root}}}\tilde{H}'.\nonumber
\end{eqnarray}
By using this identification we simply put
\begin{eqnarray}
\Lambda_{T',\varphi}\star U^t_{T,\varphi}(\vec{\tau})\nonumber\\
=U^t_{T\setminus T',\varphi}(\vec{\tau})\Lambda_{T',\varphi}.\nonumber
\end{eqnarray}
Now let us define the renormalized amplitudes, by means of the
counterterms \(\Lambda_T\) by the following formula:
\begin{eqnarray}
(R_\Lambda U)(t,-\infty) \nonumber\\
=\sum \limits_T \frac{\lambda^{n_T}}{n_T!}\sum
\limits_{T'\subseteq T} \int \Lambda_{T'}\star U_T^t(\vec{\tau})
d\vec{\tau}.\nonumber
\end{eqnarray}

d) The renormalized amplitudes \((R_\Lambda U)(t,-\infty)\rangle\) are
finite.

e) Let \(T\) be an arbitrary connected tree without shoots.
Consider the following element of \(\mathcal{H}'\):
\begin{eqnarray}
a:=\sum \limits_T \sum \limits_{T'\subseteq T} \int
\Lambda_{T'}\star U^t_T(\vec{\tau})\rangle d \vec{\tau}.\nonumber
\end{eqnarray}
We can represent the element \(a\) as follows:
\begin{eqnarray}
a=\sum \limits_{k,l,f,g=0}^{\infty} \int
w_m(x_1,...,x_{k}|y_1,...,y_{l}|v_1,...,v_{f}|w_1,...,w_{g})\nonumber\\
:\prod \limits_{i=1}^{k_m} a_+^+(x_i)dx_i \prod
\limits_{i=1}^{l_m} a_+(y_i)d y_i \prod \limits_{i=1}^{f_m}
a_-(v_i)d v_i\prod \limits_{i=1}^{g_m} a_-^+(w_i) d w_i:\rangle.\nonumber
\end{eqnarray}
Let \(\tilde{w}_{k,l,f,g}(z_1,...,z_n)\) (\(n=k_m+l_m+f_m+g_m\))
be a Fourier transform of
\(w_{k,l,f,g}(x_1,...,x_{k}|y_1,...,y_{l}|v_1,...,v_{f}|w_1,...,w_{g})\).
Then
\begin{eqnarray}
 \int dz_1,...,dz_n
\tilde{w}_{k,l,f,g}(z_1+s(1)e_1a,...,z_n+s(n)e_1a)f(z_1,...,z_n),\nonumber
\end{eqnarray}
tends to zero as \(a\) as \(a\rightarrow +\infty\). Here \(s(i)\)
are the numbers from \(\{0,1\}\) and there exist numbers
\(i,j,\;i,j=1,...,n\) such that \(s(i)=0,\;s(j)=1\) for some
\(i,j=1,...,n\). \(f(z_1,...,z_n)\) is a test function. \(e_1\) is
a unit vector parallel to the \(x\)-axis.

\textbf{Remark.} The property d) implies the weak cluster property
of the functional \((RU)(0,-\infty)\).

This theorem is a simple consequence of the theorem-construction from the section 8.

The renormalized amplitudes satisfy to the following properties:

\textbf{Property 1.} For each \(t \in \mathbb{R}\)
\begin{eqnarray}
(R_\Lambda U)(t,-\infty)\rangle\nonumber\\
=e^{-\mathcal{L}_0 t}(R_\Lambda U)(0,-\infty)\rangle.\nonumber
\end{eqnarray}
This property simply follows from the definition of \((R_\Lambda
U)(t,-\infty)\rangle\) and means that the functional \((R_\Lambda
U)(t,-\infty)\rangle\) is a stationary state.

\textbf{Property 2.}

\begin{eqnarray}
(R_\Lambda U)(t,-\infty)\rangle=U(t,0)(R_\Lambda
U)(0,-\infty)\rangle.\nonumber
\end{eqnarray}
This property follows from the following representation of \(
(R_\Lambda U)(t,-\infty)\rangle\).
\begin{eqnarray}
(R_\Lambda U)(t,-\infty)\rangle= U(t,-\infty)\mathcal{I}\rangle,\nonumber
\end{eqnarray}
where
\begin{eqnarray}
\mathcal{I}\rangle= \sum \limits_{T}
\frac{1}{n_T!}\Lambda_T\rangle,\nonumber
\end{eqnarray}
and the sum in the last formula is taken over all von Neumann
trees without shoots. Property 2 means that the functional
\((R_\Lambda U)(t,-\infty)\rangle\) satisfies to the von Neumann
dynamics.

\textbf{Remark.} The existence of the stationary translation
invariant functional satisfying to the weak cluster property
follows from the previous theorem and the properties 1, 2.

\section{The Friedrichs diagrams}

Now let us start to give a constructive description of the
counterterms \(\Lambda_T\) such that the amplitude
\(R(U)(t,-\infty)\rangle\) is finite, and the counterterms
\(\Lambda_T\) satisfy the properties a) --- e) from the previous
section.

At first we represent \(U^t_{T,\varphi}(\vec{\tau})\), where \(T\)
is some tree without shoots, as a sum taken over all so-called
Friedrichs graphs \(\Phi\) concerned with \(T\).

\textbf{Definition.} A Friedrichs graph \(\Phi_T\) concerned with
 the directed tree \(T\) without shoots is a set
 \((\tilde{V},R,Or,f^+,f^-,g)\), where \(\tilde{V}\) is a union of the set of vertices of
 \(T\) and the set \(\{\oplus\}\). Recall that there is a partial
 order on \(V_T\). We define a partial order on the set \(\tilde{V}\) if
 we put \(\forall v \in V_T\) \(\oplus> v\). \(f^+\) and \(f^-\)
 are the maps \(f^+,f^-:R\rightarrow V\) such that \(f^+(r)>f^{-}(r)\).
 \(Or\) is a map \(R\rightarrow \{+,-\}\) called an orientation.
 \(g\) is a function which to each pair \((v,r)\), \(v \in V_T\),
 \(r \in R\) such that \(f^+(r)=v\) or \(f^-(r)=v\) assigns \(+\)
 or \(-\). The graph \((\tilde{V},R,Or,f^+,f^-,g)\) must satisfy the
 property: if we consider \(\oplus\) as a vertex, the obtained
 graph is connected.

 If \(f^+(r)=v\) we write \(r\rightarrow v\), and if \(f^-(r)=v\) we write \(r\leftarrow
 v\).

If we want to point out that the object \(B\) concerned with the
graph \(\Phi\) we will write \(B_\Phi\). For example we will write
\(V_\Phi\) and \(R_\Phi\) for the sets of vertices and lines of
\(\Phi\) respectively.

 At the picture we will represent the elements of \(V\) by points and
 the element \(\oplus\) by \(\oplus\). We will represent the elements
 of \(R\) by lines. The line \(r\) connects the vertices
 \(f^+(r)\) and \(f^{-}(r)\) at the picture. We will represent
 orientation \(Or(r)\) by arrow on \(r\). If \(Or(r)=+\) the
 arrow is directed from \(f^-(r)\) to \(f^+(r)\). If \(Or(r)=-\) the
 arrow is directed from \(f^+(r)\) to \(f^-(r)\). To represent the
 map \(g:(r,v)\rightarrow \{+,-\}\) we will draw the symbol
\(g((r,v))\) (\(+\)
 or \(-\)) near each shoot \((r,v)\). At the picture a shoot \((r,v)\) is a
 small segment of the line \(r\) near \(v\).

 \textbf{Definition.} The Friedrichs diagram \(\Gamma\) is a set
 \((T,\Phi,\varphi,h)\), where \(T\) is a tree, \(\Phi\) is a
 Friedrichs graph, \(\varphi\) is a map which assigns to each
 vertex \(v\) of \(T\) a function of momenta \(\{p_r|
 r \in R_\Phi \}\) of the form
 \begin{eqnarray}
\varphi_v(...p_{r\leftrightarrows
v}...)=\psi_v(...p_{r\leftrightarrows v}...)\prod
\limits_{S_i}\delta(\sum \limits_{j=1}^{j_i} \pm p_i^j),\nonumber
 \end{eqnarray}
where \(\psi_v\) is a test function of momenta coming into (from)
the vertex \(v\). \(\{S_i\}_{i=1}^{n_v}\) is a decomposition of
the set of shoots of \(v\) into \(n_v\) of disjunctive nonempty sets
\(S_i\), \(p_i^1,...,p_i^{j_i}\) are momenta corresponding to the
shoots from \(\{S_i\}\), \(h\) is a function which assigns to each
pair \(v \in V\), \(r \in R\) such that \(f^+(r)\geq v \geq f_-(r)\) a real
positive number \(h(v,r)\).

It will be clear that it is enough to consider only the diagrams \(\Gamma\) such that
for each its vertex \(v\) and set \(S_i\in \{S_i\}_{i=1}^{n_v}\) there exists a line \(r\) such that
\((r,f^{-}(r))\in S_i\).

To each Friedrichs diagram \(\Gamma=(T,\Phi,\varphi)\) we assign
 an element of \(\tilde{\mathcal{H}}''_c\) of the form
 \begin{eqnarray}
 U^t_{(T,\Phi,\varphi)}(\vec{\tau})\nonumber\\
 =\int ...dp_{ext}...U^t_\Gamma(...p_{r_{ext}}...)\nonumber\\
 \times :...a^{\pm}_{\pm}(p_{r_{ext}})...:\rangle.\nonumber
 \end{eqnarray}
 Here \(p_{r_{ext}}\) are momenta of external lines, i.e. such lines
 \(r\) that \(f^+(r)=\oplus\). We choose the lower index of
 \(a^{\pm}_{\pm}(p_{r_{ext}})\)
 by the following rule.
 Let \(v\) be a vertex such that \(f^-(r_{ext})=v\). If
 \(g((r,v))=+\) we choose \(+\) as a lower index, and if
\(g((r,v))=-\) we choose \(-\) as a lower index. We choose the upper
index of \(a^{\pm}_{\pm}(p_{r_{ext}})\) by the following rule. If
the lower index of \(a^{\pm}_{\pm}(p_{r_{ext}})\) is \{-\} then
the upper index is equal \(+\) if the corresponding line comes
from the vertex \(v\) and this index is equal \(-\) if the
corresponding line comes into the vertex \(v\). If the lower index
of \(a^{\pm}_{\pm}(p_{r_{ext}})\) is \{+\} then the upper index is
equal \(-\) if the corresponding line comes from the vertex \(v\)
and this index is equal \(+\) if the corresponding line comes into
the vertex \(v\).

Now let us describe the amplitude \(U_\Gamma^t(...p_{ext}...)\).
By definition we have
\begin{eqnarray}
U^0_\Gamma(\vec{\tau})(...p_{ext}...)\nonumber\\
=\int \limits_{r \in R_{in}} \prod \limits_{v} \varphi_v(...
p_{r\leftrightarrows v}...)  \nonumber\\
\times \prod \limits_{r \in R_\Gamma} e^{i Or(r) p_r^2 (\sum
\limits_{r_T \in (R_T)_r} \tau_{r_T}+\sum \limits_{v \in
V_r}h(v,r))}dp_r \nonumber\\
\times\prod \limits_{r \in R} G(Or(r), g((r,f^+(r))),g((r,f^-(r))))(p).
\nonumber
\end{eqnarray}

Let us describe the elements of this formula. \(R_\Gamma\) is a set of
all lines of diagram \(\Gamma\). Symbol \(r\leftrightarrows v\)
denotes that the line \(r\) comes into (from) the vertex \(v\). In
the expression
\begin{eqnarray}
\psi_v(... p_{r\leftrightarrows v}...)\delta(\sum
\limits_{r\leftrightarrows v} \pm p_r)\nonumber
\end{eqnarray}
we take the upper sign \(+\) if the line \(r\) comes into the
vertex \(v\) and we take lower sign \(-\) in the opposite case.
The symbol \(R_T\) denotes the set of lines of the tree \(T\) from
the triple \((T,\Phi,\varphi)\) and symbol \(r_T\) means the line
from \(R_T\). The symbol \(V_r\) denotes the set of all vertices
\(v\) such that \(f^+(r)\geq v\geq f^-(r)\). The symbol \((R_T)_r\)
denotes the set of all lines \(r_T\) of \(R_T\) such that the
increasing path coming from \(f^-(r)\) into \(f^+(r)\) contains
\(r_T\). \(G(Or(r),g(f^+(r)),g(f^-(r)))(p)\) is a factor defined
as follows
\begin{eqnarray}
G(Or(r),g(f^+(r)),g(f^-(r)))(p)\delta(p-p')\nonumber\\
=\rho_0'(a^{sgn(-Or(r)g((r,f^+(r))))}_{g((r,f^+(r)))}(p),a^{sgn(Or(r)g((r,f^-(r))))}_{g((r,f^-(r)))}(p')).\nonumber
\end{eqnarray}
Below we will simply write \(G_r(p)\) instead of
\(G(Or(r),g(f^+(r)),g(f^-(r)))(p)\).

It is evident that we can represent \(U^0_T(\vec{\tau})\) as a sum
taken over some Friedrichs diagrams \(\Gamma\) corresponding to the tree
\(T\) of the quantities \(U^0_\Gamma (\vec{s})\).

Now let us define the quotient diagrams.

\textbf{Definition}. Let \(\Gamma=(T,\Phi,\varphi,h)\) be a
Friedrichs diagram and \(A\subset R_T\) be a subset of the set
\(R_T\) of lines of \(T\) and \(\vec{\tau}\) be a map from \(R_T\)
into \(\mathbb{R}^+\).

We define the quotient diagram
\(\Gamma_{A\vec{\tau}}:=(T_A,\Phi_A,\varphi_{A\vec{\tau}},h_A)\)
in the following way. To obtain the tree \(T_A\) we must tighten
all lines from \(A\) into points. To obtain \(\Phi_A\) we must
remove all loops obtained by tightening all lines from \(A\) into
the point.

Now let us define \(\varphi_{A\vec{\tau}}\). Joint all the
vertices of \(T\) to \(A\). We obtain a tree denoted by \({}^AT\).
Let \(\{C{}^AT\}\) be a set of all connected components of
\({}^AT\). Let \(v_0\) be a vertex of \(\Phi_A\) corresponding to
the connected component \(C{}^AT\) of \({}^AT\). Put by
definition:
\begin{eqnarray}
\varphi_\Gamma(...p_{r\leftrightarrows
v}...)_{A\vec{\tau}}\nonumber\\
=\int  \prod \limits_{v \in V}\varphi_v(...\pm
p_{r\leftrightarrows v}...) \nonumber\\
\times \prod \limits_{r \in R_{in}} e^{iOr(r)p_r^2(\sum \limits_{r
\in (R_T)_r} \tau_{r_T}+\sum \limits_{v \in V_r} h(v,r))}.\nonumber
\end{eqnarray}

Let us point out the notations in the previous formula. \(R_{in}\)
is a set of all lines of \(\Phi_A\) such that \(f^+(r)\) and
\(f^-(r)\) are the vertices of \(C{}^AT\). \((R_T)_r\) denotes the
set of all lines \(r_T\) of \(R_T\) such that the increasing path
coming from \(f^-(r)\) into \(f^+(r)\) contains \(r_T\). The
symbol \(V_r\) denotes the set of all vertices \(v\) such that
\(f^+(r)\geq v\geq f^-(r)\). \(h_A(v_0,r)=\sum \limits_{v \in
V_{C{}^AT}}h(v,r)+\sum \limits_{r_T \in A;\; r_T \in (R_T)_r}
\tau_{r_T}\).

\textbf{Definition.} Let \(\Gamma\) be a Friedrichs diagram. Let
\(\mathcal{F}_\Gamma\) be a space of all functions of external
momenta of the diagram \(\Gamma\) of the form:
\begin{eqnarray}
 \psi(...p_{ext}...),\nonumber
\end{eqnarray}
where \(\psi(...p_{ext}...)\) is a test function of external momenta.

The convolution of the amplitude
\(A_\Gamma(\vec{\tau})(...p_{ext}...)\) with the function \(f \in
\mathcal{F}_\Gamma\) we denote by \(A_\Gamma(\vec{\tau})[f]\).

\section{The Bogoliubov --- Parasiuk renormalization
prescriptions}

Let for each Friedrichs diagram \(\Gamma=(T,\Phi,\varphi)\)
\(A_\Gamma(\vec{\tau})(...p_{ext}...)\) be some amplitude. Fix
some diagram \(\Gamma\) and let \(T'\) be some right subtree of
the tree \(T\) corresponding to \(\Gamma\). Let \(\Gamma_{T'}\) be
a restriction of the diagram \(\Gamma\) on \(T'\) in obvious
sense. Define the amplitude \(A_{\Gamma_{T'}}\star U_\Gamma
(...p_{ext}...)\) by the following formula:
\begin{eqnarray}
A_{\Gamma_{T'}}\star U_\Gamma (...p_{ext}...)\nonumber\\
=\int \prod \limits_{r \in R'}\{e^{iOr(r)p_r^2(\sum \limits_{r_T \in
(R'_T)_r}\tau_{r_T}+\sum \limits_{v \in V'_r}h(v,r))}\}\prod
\limits_{v \in
V'}\varphi_v(...p_{r\rightleftarrows v}...)\nonumber\\
\times A_{\Gamma_{T'}}(...p...).\nonumber
\end{eqnarray}
In this formula \(V'\) is a set of all vertices \(v\) such that
\(v\) is not a vertex of \(V_{T'}\), \(R'\) is a set of all lines
\(r\) of \(\Phi_\Gamma\) such that \(f^+(r)\) is not a vertex of
\(T'\).
\((R_T')_r\) is a set of all lines \(r_T\) of \(T\) such that
\(r_T\) is not a line of \(T'\) and there exists an increasing
path on \(T\) coming from \(f^{-}(r)\) into \(f^+(r)\) such that
this path contains \(r_T\). \(V'_r\) is a set of all vertices
\(v\) of T such that \(v\) is not a vertex of \(T'\) and
\(f^+(r)\geq v\geq f^-(r)\).

Let \(A_\Gamma(\vec{\tau})(p)\) be some amplitude. Put by
definition:
\begin{eqnarray}
\hat{A}_\Gamma(s_1,...,s_n)(p):=A_\Gamma(\frac{1}{s_1},...,\frac{1}{s_n})(p)\prod
\limits_{i=1}^{n} \frac{1}{s_i^2},\nonumber
\end{eqnarray}
where \(n\) is a number of lines of \(T_\Gamma\). Below we will
consider the amplitudes \(\hat{A}_\Gamma(\vec{s})[f]\) as
distributions on \((\mathbb{R}^+)^n\) i.e. as an element of the
space of tempered distributions \(S'((\mathbb{R}^+)^n)\). Let
\(\psi(\vec{s})\) be a test function from \(S((\mathbb{R}^+)^n)\).
The convolution of the amplitude \(\hat{A}_\Gamma(\vec{s})[f]\)
and the function \(\psi(\vec{s})\) we denote by:
\begin{eqnarray}
\langle\hat{A}_\Gamma(\vec{s})[f],\psi(\vec{s})\rangle\nonumber\\
:=\int \limits_{(\mathbb{R}^+)^n} d\vec{s}
\hat{A}_\Gamma(\vec{s})[f]\psi(\vec{s}).\nonumber
\end{eqnarray}

\textbf{The Bogoliubov --- Parasiuk prescriptions.} It will be clear below,
that we can take into account only the diagrams \(\Gamma\) such that for each line \(r_T\)
of the corresponding tree of correlations \(T\) \(\sharp R_{r_T}\geq 3\). Here \(R_{r_T}\) is a set of all lines \(r\) of
\(\Gamma\) such that the increasing path on \(T\) which connects \(f^-(r)\) and \(f^+(r)\) contains \(r_T\).
Below we will consider only such diagrams. Other diagram can be simply subtracted by some counterterms
\(\Lambda_T\).

According to the Bogoliubov --- Parasiuk prescriptions we must to
each diagram \(\Gamma\) (corresponding to the connected tree)
assign the counterterm amplitude \(\hat{C}_\Gamma(\vec{s})[f]\)
\(f \in \mathcal{F}_\Gamma\) satisfying the following properties.

a) (Locality.) \(\hat{C}_\Gamma(\vec{s})[f]\) is a finite linear
combination of \(\delta\) functions centered at zero and their
derivatives.

b) Let \(\Gamma\) be a Friedrichs diagram and \(T\) be a
corresponding tree of correlations. Let \(A \subseteq R_T\) and
\(T'\) is some right subtree of \(T\) such that:

1) all lines \(r_T\) of T such that \(r_T\) is not a line of
\(R_{T'}\) belong to \(A\),

2) All the root lines of \(T'\) do not belongs to \(A\).

Then
\begin{eqnarray}
\hat{C}_{\Gamma_{A\vec{\tau}}}(\vec{s})[f]=(\hat{C}_{\Gamma_{A'\vec{\tau}}'}\star
\hat{U}_\Gamma)(\vec{s})[f],\nonumber
\end{eqnarray}
where \(A':=A\cap (R_{T'})\) and \(\Gamma'\) is a restriction of
\(\Gamma\) on \(T'\).

c)
\begin{eqnarray}
\hat{C}_\Gamma(\vec{s})[f]=-\mathbb{T}(\sum
\limits_{\emptyset\subset A\subset R_{T_\Gamma}}
\hat{C}_{\Gamma_{A\vec{\tau}}}(\vec{s})[f]+\hat{U}_{\Gamma}(\vec{s})[f]),\nonumber
\end{eqnarray}
where
\(\vec{\tau}=(\tau_1,...,\tau_n)=(\frac{1}{s_1},...,\frac{1}{s_n})\),
the symbol \(\subset\) means here the strong inclusion and
\(\mathbb{T}\) is some subtract operator.

d) The amplitudes \(\hat{C}_\Gamma(\vec{s})[f]\) satisfy to the property
of time-translation invariance, i.e.
\begin{eqnarray}
e^{i\sum \limits_{r \in (R_{root})_\Gamma} Or(r) p_r^2 t}
\hat{C}_\Gamma(\tau_1,...,\tau_n)[f] =\hat{C}_\Gamma(\tau_1+t,...,\tau_n)[f].\nonumber
\end{eqnarray}

e) Let \(\Gamma\) be a Friedrichs diagram. Let
\begin{eqnarray}
\hat{R}_\Gamma'(\vec{s})[f]:=\hat{U}_\Gamma(\vec{s})[f]+\sum
\limits_{\emptyset \subset A\subset R_{T_\Gamma}}
\hat{C}_{\Gamma_{A\vec{\tau}}}(\vec{s})[f],\; \rm and \mit \nonumber\\
\hat{R}_\Gamma(\vec{s})[f]:=\hat{U}_\Gamma(\vec{s})[f]+\sum
\limits_{\emptyset \subset A\subset R_{T_\Gamma}}
\hat{C}_{\Gamma_{A\vec{\tau}}}(\vec{s})[f]+\hat{C}_\Gamma(\vec{s})[f].\nonumber
\end{eqnarray}
The amplitudes \(\hat{R}_\Gamma(\vec{s})\) are well defined
distributions on \((\mathbb{R}^+)^n\).

f) The amplitudes \(\hat{R}_\Gamma(\vec{s})\) satisfy the weak
cluster property. This property means the following. Let
\(f(...p_{ext}...)\) be a test function. Then
\begin{eqnarray}
\int dp \hat{R}_\Gamma(\vec{s})(...p_{ext}...) f(...p_{ext}...)
e^{ia\sum \limits_{r \in A} p_r^1}\rightarrow 0,\nonumber
\end{eqnarray}
as \(a\rightarrow \infty\). Here \(p_r^1\) is a projection of \(p_r\) to the \(x\)-axis.

Put by definition for each diagram \(\Gamma\):
\begin{eqnarray}
\Lambda_\Gamma(\vec{\tau})=\sum \limits_{A\subseteq R_{T_\Gamma}}'
C_{\Gamma_{A\vec{\tau}}},\nonumber
\end{eqnarray}
where \('\) in the sum means that all the root lines of
\(T_\Gamma\) do not belong to \(A\).

Put
\begin{eqnarray}
\Lambda_T=\sum \limits_{\Gamma\sim T}
\int
\limits_{(\mathbb{R}^+)^n}d\vec{\tau}\Lambda_\Gamma(\vec{\tau})(...p_{ext}...)...a^\pm_\pm(p_{ext})...\rangle,\nonumber
\end{eqnarray}
where the symbol \(\Gamma\sim T\) means that the sum is taken over all
diagrams corresponding to \(T\) with suitable combinatoric
factors. Suppose that the properties a) --- f) are satisfied. Then
\(\Lambda_T\) are the counterterms needed in the section 8. Not that the state corresponding to
\((RU)(t,-\infty)\rangle\) will obvious commute with the number of particle operator.

\textbf{Theorem --- Construction.} \textit{It is possible to find such a
subtract operator \(\mathbb{T}\) such that there exist
counterterms \(\hat{C}_\Gamma\) satisfying the properties a)
--- f).}

Note that it is not necessarily for us use not real counterterms. Indeed the evolution
operator is real, so after renormalization we can simply take
\(\rm Re \mit\; (RU)(t,-\infty)\rangle\).

\section{Proof of the theorem-construction}

In this section we prove the theorem-construction from the previous section. Note that to
prove our theorem we will use some ideas of the papers \cite{10,11,12}.

Before we prove our theorem let us prove the following

\textbf{Lemma 1.}\textit{ Let \(L_1=S(\mathbb{R}^{k})\),
\(L_2=S((\mathbb{R}^+)^n)\), \(k,n=1,2,...\). Let \(A(p)\) be some
nonzero quadratic form on \(\mathbb{R}^{k}\). Let \(T^1_t\),
\(t\geq 0\) be an one-parameter semigroup acting in \(L_1\)
drfined as follows:}
\begin{eqnarray}
T^1_t:f(...p...)\mapsto e^{iA(p)t}f(...p...).\nonumber
\end{eqnarray}

\textit{Let \(T^2_t\) \(t\geq 0\) be some infinitely differentiable
semigroup of continuous operators in} \(L_2\).

\textit{Let \(M\) be a subspace of finite codimension in \(L_2\). Suppose
that \(M\) is invariant under the action of} \(T^2_t\), i.e.
\(\forall t>0\) \(T^2_t M\subset M\).

\textit{Suppose that there exist the linear independent vectors \(f_1,...f_l\) in \(L_2\)
such that}

\begin{eqnarray}
\rm Lin \mit \{\{f_1,...,f_l\},
M\}=L_2,\nonumber\\
 M\cap \rm Lin \mit \{f_1,...,f_l\}=0.\nonumber
 \end{eqnarray}
and for each \(i=1,...,l\), \(t\geq 0\)
\begin{eqnarray}
T_t^2 f_i=f_i+a_{i-1}f_{i-1}+...+a_1 f_1+f,\nonumber
\end{eqnarray}
\textit{for some coefficients \(a_{i-1},...,a_1\) and the element} \(f \in
M\).

\textit{Let \(g\) be a functional on \(L_1\otimes M\) such that \(g\) is
continuous with respect to the topology on \(S(\mathbb{R}^k)\times
S((\mathbb{R}^+)^n)\). Suppose that \(\forall f \in L_1\otimes M\)
and \(\forall t>0\) \(\langle g, T_t f\rangle=\langle g,f
\rangle\) where} \(T_t=T^1_t\otimes T^2_t\).

\textit{Then, there exists an continuous extension \(\tilde{g}\) of \(g\)
on \(S(\mathbb{R}^k)\times S((\mathbb{R}^+)^n)\) such that
\(\forall f \in L_1\otimes L_2\) and \(t>0\) \(\langle
\tilde{g},T_t f\rangle=\langle\tilde{g},f\rangle\).
}

By definition we say that the functional \(h\) on \(L_1\otimes
L_2\) is invariant if \(\forall t>0\) and \(\forall f \in
L_1\otimes L_2\) \(\langle h,T_t f\rangle=\langle h,f\rangle\).

\textbf{Proof of the lemma 1.} At first we extend our functional
\(g\) to the invariant functional \(\tilde{g}\) on \(L_1\otimes
L_2\) and then we prove that \(\tilde{g}\) is continuous.

Let \(N\) be a subspace of \(L_1\) of all functions of the form,
\(A(p)f(p)\), where \(f(p)\) is a test function. Let \(M_1=\rm Lin
\mit \{M\cup \{f_1\}\}\). Let

\begin{eqnarray}
h:=(\frac{d}{dt}T^2_t)|_{t=0} f_1.\nonumber
\end{eqnarray}
Let \(k\) be a continuous functional on \(N\) defined as follows:
\begin{eqnarray}
\langle k,\varphi(p)A(p)\rangle=-\langle g,\varphi(p)\otimes\nonumber
h\rangle. \label{ZUZU}
\end{eqnarray}
Let \(\tilde{k}\) be an arbitrary continuous extension of \(k\) on
whole space \(L_1\). The existence of such continuation follows from Malgrange's preparation theorem \cite{13}.
 Now we define the continuous functional
\(\tilde{g}_1\) on \(L_1\otimes M_1\) as follows:
\begin{eqnarray}
\tilde{g}_1|_{L_1\otimes M}=g|_{L_1\otimes M},\nonumber\\
\langle \tilde{g}_1,f\otimes
f_1\rangle=\langle\tilde{k},f\rangle,\; \forall f \in L_1.\nonumber
 \end{eqnarray}

 According to (\ref{ZUZU}) we find that \(\tilde{g}_1\) is an
 invariant extension of \(g\) on \(L_1\otimes M_1\). Step by step we can extend by the same procedure
 the functional \(g\) to
 the functionals \(\tilde{g}_2\),
 ....\(\tilde{g}_l\) on \(L_1\otimes M_1\),...,\(L_l\otimes M_l\)
 respectively, where
 \( M_2= \rm Lin
\mit \{M\cup \{f_1,f_2\}\}\),...,\(M_l=\rm Lin \mit \{M\cup
\{f_1,f_2,...,f_l\}\}\) respectively. Just constructed functional
is separately continuous so it is continuous. The lemma is proved.

\textbf{Sketch of the proof of the theorem.} We will prove the theorem by
induction on the number of lines of the tree of correlations
\(T_\Gamma\) corresponding to the diagram \(\Gamma\). It is evident that it is enough to
consider only the diagrams with connected tree of correlations.

The base of induction is evident. Suppose that the theorem is
proved for all diagrams of order \(<n\). (Order is a number of
lines of the tree of correlations.)

Let us give some definitions. Let \(\xi(t)\) be a smooth function
on \([0,+\infty)\) such that \(0\leq \xi(t)\leq 1\), \(\xi(t)=1\)
in some small neigborhood of zero and \(\xi(t)=0\) if
\(t>\frac{1}{3n}\). Let us define a decomposition of unit
\(\{\eta_A(\vec{s})|A\subset \{1,...,n\}\}\) by the formula
\begin{eqnarray}
\eta_A(\vec{s})=\prod \limits_{i \notin A} \xi(s_i) \prod
\limits_{i \in A}(1-\xi(s_i)).\nonumber
\end{eqnarray}

Let \(\psi(x)\) be some test function on real
line such that \(\psi(t)\geq 0\), \(\int \psi(t)dt=1\) and
\(\psi(t)=0\) if \(|t|>\frac{1}{10}\). Put by definition:
\begin{eqnarray}
\delta_\lambda(x-\lambda)=\frac{x}{\lambda^2}\psi(\frac{x-\lambda}{\lambda}).\nonumber
\end{eqnarray}
We have
\begin{eqnarray}
\int \limits_0^{+\infty}d \lambda \delta_\lambda(x-\lambda)=1.\nonumber
\end{eqnarray}
Let \(S_N((\mathbb{R}^+)^n)\), \(N=1,2,...,\) be a subspace of
\(S((\mathbb{R}^+)^n)\) of all functions \(f\) such that \(f\) has
a zero of order \(\geq N\) at zero. Let \(\Psi(\vec{s})\) be a function of \(S((\mathbb{R}^+)^n)\).

 We have:
\begin{eqnarray}
\langle\hat{R}_{\Gamma}(\vec{s})[f],\Psi(\vec{s}) \rangle \nonumber\\
=\sum \limits_{A\subset \{1,...,n\}} \int \limits_0^{+\infty}d
\lambda \lambda^{n-1} \int \limits_{(\mathbb{R}^+)^n} d\vec{s}
\hat{R}_{\Gamma_{A \lambda
\vec{s}}}(\vec{s}|_{\{1,...n\}\setminus A} )\nonumber\\
\delta_1(1-|\vec{s}|)\Psi(\lambda\vec{s})\eta_A(\vec{s}) .\nonumber
\label{GF}
\end{eqnarray}

The inner integral in (\ref{GF}) converges according to the
inductive assumption. Therefore if \(\Psi(\vec{s}) \in
S_N((\mathbb{R}^+)^n)\) and \(N\) is large enough the integral at
the right hand side of (\ref{GF}) converges. So
\begin{eqnarray}
\langle\hat{R}_{\Gamma}(\vec{s} )[f],\Psi(\vec{s}) \rangle\nonumber
\end{eqnarray}
defines a separately continuous functional on
\(S(\mathbb{R}^{3f})\otimes S_N((\mathbb{R}_+)^n)\). \(f=l-1\), where \(l\) is a number of
 external lines of \(\Gamma\). To define a subtract
operator \(\mathbb{T}\) we must extend the functional
\(\langle\hat{R}_{\Gamma}(\vec{s})[f],\Psi(\vec{s}) \rangle\) to
the space \(S(\mathbb{R}^{3f})\otimes S((\mathbb{R})^n)\) such
that extended functional will satisfy to time-translation
invariant property. To obtain this extension we use the lemma. In
our case \(L_1=S(\mathbb{R}^{3f})\), \(L_2=S((\mathbb{R}^+)^n)\),
\(A(p)=-\sum \limits_{r \in R_{ext}} Or(r)p_r^2\). \(T_t^2\) is an
operator acting in the \(S((\mathbb{R}_+)^n)\) as follows.
\begin{eqnarray}
T_t^2 f(s_1,...,s_n)=f(\frac{s_1}{1-s_1t},s_2,...,s_n)\; \rm if
\mit\; s_i<\frac{1}{t},\nonumber\\
T_t^2 f(s_1,...,s_n)=0,\; \rm if \mit\; s_i\geq \frac{1}{t}.\nonumber
\end{eqnarray}
The basis \(\{f_1,...f_l\}\) from the lemma is
\(\{s_1^{m_1}....s^{m_n}_n\eta_\emptyset(\vec{s})\}\),
\(m_1,...m_n=1,2,3...\), \(m_1+m_2+...+m_n\leq N\)
lexicographically ordered. We can now apply our lemma directly.

Now let us prove the weak cluster property. Let \(p \in\mathbf{R}^3\). Denote by
\(p^1,\;p^2,\;p^3\) the projections of \(p\) to the \(x,\;y,\;z\)-axis respectively. To prove the weak
cluster property it is enough to prove the following statement: for each connected diagram \(\Gamma\)
the function \(\langle F_\Gamma(\vec{s})(...p_{ext}...),\Psi(\vec{s})\rangle\) defined by
\begin{eqnarray}
\delta(\sum \pm p_{ext})\langle F_\Gamma(\vec{s})(...p_{ext}...),\Psi(\vec{s})\rangle=
\langle\hat{R}_\Gamma(\vec{s})(...p_{ext}...),\Psi(\vec{s})\rangle\nonumber
\end{eqnarray}
is a distribution of variables \(...p_{ext}^2...p_{ext}^3...\)
(constrained by momentum conservation law) which depends on
\(...p_{ext}^1...\) (constrained by momentum conservation law) by
the continuously differentiable way. We will prove this statement
by induction on the number of lines of the corresponding tree of
correlations. The base of induction is evident. Suppose that the
statement is proved for all the trees of correlations such that
the number of their lines \(<n\). Let \(\Gamma\) be a diagram such
that the number of the lines of the corresponding tree of
correlations is equal to \(n\). It is evident that if
\(\Psi(\vec{s})\) has a zero of enough high order at zero then
\(\langle\hat{F}_\Gamma(\vec{s})(...p_{ext}...),\Psi(\vec{s})\rangle\)
belongs to the required class (its enough to use our construction
with decomposition of unit). Therefore we need to solve by
induction the system of equations of the form:
\begin{eqnarray}
(i\sum \pm (p_{ext})^2)\langle F_\Gamma(\vec{s})(...p_{ext}...),\Psi(\vec{s})\rangle\nonumber\\
=\langle
F_\Gamma(\vec{s})(...p_{ext}...),\frac{d}{dt}T_t^{2}\Psi(\vec{s})\rangle.\nonumber
\end{eqnarray}
According to Malgrange's preparation theorem \cite{13} we can
choose the solution \(\langle
F_\Gamma(\vec{s})(...p_{ext}...),\Psi(\vec{s})\rangle\) such that
it belongs to the required class if \( \langle
F_\Gamma(\vec{s})(...p_{ext}...),\frac{d}{dt}T_t^{2}\Psi(\vec{s})\rangle\)
belongs to the required class. Therefore the statement is proved.
So our theorem is proved.
\section{Derivation of non ergodic property from main result}
 Let us prove (more accurately as in
introduction) that our system (Bose gas with weak pair interaction
in thermodynamical limit) is non-ergodic system.

Let us recall definition of ergodicity \cite{45}.

\textbf{Definition.} Consider a quantum system described by
Hamiltonian \(H\). This system is said to be ergodic if the
spectrum of \(H\) is simple.

This definition is equivalent to the following

\textbf{Definition.} A quantum system described by Hamiltonian
\(H\) is said to be ergodic if each bounded operator commuting
with \(H\) is a function of \(H\).

The generalization of this definition to the case when where
exists some additional commuting first integrals is obvious
\cite{45}.

It is to difficult to define a Hilbert space and Hamiltonian (as a
self-adjoint operator in Hilbert space of states) of the system,
in thermodynamical limit. So we give some new definition of
ergodicity for this case which can be considered as some variant
of last definition. Let us introduce some useful notations. Let
\(V\) be an algebra of all Wick monomials with kernels from the
Schwartz space, i.e \(V\) is a linear space of all expressions of
the form
\begin{eqnarray}
\int  w(p_1,...,p_n|q_1,...,q_m)\prod \limits_{i=1}^n a^+(p_i)dp_i
\prod \limits_{j=1}^m
a(q_i)dq_i,\nonumber\\
w \in S(\mathbb{R}^{3(n+m)}),\nonumber
\end{eqnarray}
where the multiplication is defined by canonical commutative
relations. Let \(V'\) be an algebraically dual of \(V\). We say
that the functional \(\rho \in V\) is a stationary functional if
\(\forall v \in V\) \(\rho([H,v])=0\). Here \(H\) is a Hamiltonian
of our Bose gas. We say that the functional \(\rho \in V\) is a
translation-invariant functional if \(\forall v \in V\)
\(\rho([\vec{P},v])=0\), where \(\vec{P}\) is an operator of
momentum of our system. We say that the functional \(\rho \in V\) commute
with the number of particle operator
if \(\forall v \in V\) \(\rho([N,v])=0\), where \(N\) ia a number of particle operator.
Note that \(\forall v \in V\)
\([H,v],\;[N,v],\;[\vec{P},v] \in V\). Denote by \(V_s\) the linear space
of all Wick monomials of the form \([H,v_0]+[N,v_1]+[\vec{P},\vec{v}]\),
\(v_0,v_1,\vec{v}\in V\). Denote by \(V'_s\) the space of all
translation-invariant stationary states commuting with the number of particle operator. We
have \(\forall f \in V'(f \in V'_s \Leftrightarrow\; \forall v \in V_s\: f(v)=0)\). Now
let us introduce a notion of Gibbsian states.

 Let \(\beta,\;\mu \in \mathbb{R},\:\beta>0\), \(\vec{v} \in
 \mathbb{R}^3\). We define Gibbsian state on \(V\) formally by the
 following formula:
 \begin{eqnarray}
 \langle a\rangle_{\beta,\mu,\vec{v}}=\frac{1}{Z_{\beta,\mu,\vec{v}}} {\rm tr \mit} (a
 e^{-\beta(H-\mu N+\vec{v}\vec{P})}),\nonumber
 \end{eqnarray}
 where \(a \in V\), \(N\) is a particle number operator and \(Z\) is so called
 statistical sum:
 \begin{eqnarray}
 Z_{\beta,\mu,\vec{v}}={\rm tr \mit} (e^{-\beta(H-\mu N+\vec{v}\vec{P})}).\nonumber
 \end{eqnarray}
 This states corresponds to canonical distributions. Note that one
 of the basis statement of statistical mechanics states that there
 no difference which distribution we use: canonical or
 micro-canonical distribution. Bellow we will omit \(\vec{v}\), \(\vec{P}\) at
 all formulas to simplify or notations.

 Let \(V'_G\) be a subspace spanned by all Gibbsian states, i.e.
 \(V'_G\) is a set of all functionals \(\langle\cdot\rangle\) on \(V\) of the form:
 \begin{eqnarray}
 \langle\cdot\rangle=\sum \limits_{\alpha} c_\alpha
 \langle\cdot\rangle_{\beta_\alpha,\mu_\alpha},\nonumber
 \end{eqnarray}
 where the sum is understood in some generalized sense, for
 example it may be continuous (integral). It is evident that
 \(V'_G\subseteq V'_s\).

 Now we can give the definition of ergocity for Bose gas in
 thermodynamical limit.

 \textbf{Definition.} We say that our system is ergodic if each
 translation invariant stationary state can be represented as a
 superposition of Gibbsian states, i.e \(V'_s=V'_G\).

 After these previous discussion let us start to prove our
 statement. Recall that we find non Gibbsian real stationary translation invariant
functional \(\langle\cdot\rangle\) constructed as a formal power
series on coupling constant \(\lambda\) satisfying to the weak
cluster property. This functional can be represented as follows:
\begin{eqnarray}
 \langle\cdot\rangle=\langle\cdot\rangle_0+\langle\cdot\rangle_1+...,\nonumber
\end{eqnarray}
where \(\langle\cdot\rangle_0\) is a functional of zero order of
coupling constant \(\lambda\), \(\langle\cdot\rangle_0\) is a
functional of first order of coupling constant \(\lambda\) and
e.c.t.

Suppose that our system is ergodic.

We do not suppose that the series
\(\langle\cdot\rangle_0+\langle\cdot\rangle_1+...\) converges, but
we will work with it formally as with convergent
 series and find explicit formulas for \(\langle\cdot\rangle\) under the assumption
 of ergodicity. Let us illustrate formal manipulation that we will
 use by several examples.

 \textbf{Example 1.} Let us calculate the sum \(\sum
 \limits_{i=0}^{\infty} x^i\). We do not suppose that \(|x|<1\).
 Denote by \(S\) the sum of this series. We have
 \begin{eqnarray}
 S=\sum \limits_{i=0}^{\infty} x^i=1+\sum \limits_{i=1}^{\infty} x^i=1+x\sum \limits_{i=0}^{\infty}
 x^i=1+xS.\nonumber
 \end{eqnarray}
 Therefore
 \begin{eqnarray}
 S=1+xS,\nonumber\\
 (1-x)S=1, \nonumber \\
 s=\frac{1}{1-x}.\nonumber
 \end{eqnarray}

\textbf{Example 2.} Let us calculate the sum \(\sum
\limits_{i=1}^{\infty}\frac{1}{i^2}\). Consider the function
\(\frac{\sin x}{x}=a_0x+a_1x+a_2 x^2+...\) as a polynomial of
infinite degree. Let us use the Viete theorem for this
"polynomial". The roots of \(\frac{\sin x}{x}\) are \(x_i=i
\pi,\;i \in \mathbb{Z}\setminus \{0\}\). According to the Viete
theorem we have
\begin{eqnarray}
\prod \limits_{i \in \mathbb{Z}\setminus \{0\}} x_i=C,\nonumber\\
\sum \limits_{j \in \mathbb{Z}\setminus \{0\}} \prod \limits_{i
\in \mathbb{Z}\setminus \{0,j\}}x_i=0,\nonumber
\end{eqnarray}
\begin{eqnarray}
\sum \limits_{i<j,\;i,j \in \mathbb{Z}\setminus \{0\}}\prod
\limits_{k \in \mathbb{Z}\setminus
\{0,i,j\}}x_k=-\frac{C}{6}.\nonumber
\end{eqnarray}
We find from these equations that
\begin{eqnarray}
\sum \limits_{i \in \mathbb{Z}\setminus
\{0\}}\frac{1}{x_i}=0,\nonumber
\end{eqnarray}
and
\begin{eqnarray}
\sum \limits_{i<j,\;i,j \in \mathbb{Z}\setminus
\{0\}}\frac{1}{x_ix_j}= 1/2(\sum \limits_{i \in
\mathbb{Z}\setminus \{0\}}\frac{1}{x_i})^2-1/2\sum \limits_{i \in
\mathbb{Z}\setminus \{0\}}\frac{1}{x_i^2}\nonumber\\
=-1/2\sum \limits_{i \in \mathbb{Z}\setminus
\{0\}}\frac{1}{x_i^2}=-\frac{1}{6}.\nonumber
\end{eqnarray}
Therefore
\begin{eqnarray}
\sum \limits_{i=1,2...}\frac{1}{x_i^2}=\frac{1}{6}.\nonumber
\end{eqnarray}
But \(x_n=\pi n\), so we finally have
\begin{eqnarray}
\sum \limits_{i=1}^{\infty}\frac{1}{i^2}=\frac{\pi^2}{6}.\nonumber
\end{eqnarray}
Such formal manipulation was widely used by Euler and others.
Suppose that the set of such formal rules is enough large from one
hand and does not contain a contradiction from other hand. These
rules we call the Euler rules. If we can find the "sum" of some
series by using the Euler rules then this series is called
convergent in Euler sense. The "sum" of this series is called a
sum in Euler sense.

Let us prove that our functional \(\langle\cdot\rangle\) can be
represented as follows (under the assumption of ergodicity):
\begin{eqnarray}
 \langle\cdot\rangle=\sum \limits_{\alpha} c_\alpha
 \langle\cdot\rangle_{\beta_\alpha,\mu_\alpha},\nonumber
 \end{eqnarray}
where \(c_{\alpha}\) are the "sums" of probably divergent series.
The convergence (in the Euler sense) of this series will be proven
below (under the assumption of ergodicy). The sum can be
continuous (integral).

Let \(\{e_\alpha,\alpha \in \mathfrak{A}\}\) be a Hamele basis of
\(V_s\), \(V_s={\rm Lin \mit}\{e_\alpha,\alpha \in
\mathfrak{A}\}\). Let \(\{e_\beta,\beta \in \mathfrak{B}\}\),
\(\mathfrak{A}\cap\mathfrak{B}=\emptyset\) be a completion of
\(\{e_\alpha,\alpha \in \mathfrak{A}\}\) to the Hamele basis of
\(V\), i.e. \(\{e_\alpha,\alpha \in
\mathfrak{A}\}\cup\{e_\beta,\beta \in \mathfrak{B}\}\) be a Hamele
basis of \(V\). \(\forall \gamma \in
\mathfrak{A}\cup\mathfrak{B}\) let \(f_\gamma\) be an element of
\(V'\) such that \(f_\gamma(e_\gamma)=1\) and
\(f_\gamma(e_{\gamma'})=0\) if \(\gamma\neq\gamma'\), \(\gamma'
\in \mathfrak{A}\cup\mathfrak{B}\).

An arbitrary functional \(\rho\) from \(V'\) now can be
represented as a sum
\begin{eqnarray}
\rho=\sum \limits_{\alpha \in \mathfrak{A}}l_\alpha f_\alpha +
\sum \limits_{\beta \in \mathfrak{B}}l_\beta f_\beta,\nonumber
\end{eqnarray}
where \(l_\alpha,\;l_\beta\) are arbitrary  numbers. Note that for
arbitrary \(l_\alpha,\;l_\beta\) the right hand side of last
equation is well defined because \(\forall v \in V\)
\(f_\gamma(v)\neq 0\) only for finite number of elements \(\gamma
\in \mathfrak{A}\cup\mathfrak{B}\). It is obvious now that an
arbitrary element \(f \in V'_s\subseteq V'_G\) (ergodicity) can be
represented as follows:
\begin{eqnarray}
f=\sum \limits_{\beta \in \mathfrak{B}}l_\beta f_\beta,\nonumber
\end{eqnarray}
where \(l_\beta\) are arbitrary numbers.

\(\forall i=0,1,2,...\) we have the following representations
\begin{eqnarray}
\langle\cdot\rangle_i=\sum \limits_{\alpha \in
\mathfrak{A}}s^i_\alpha f_\alpha + \sum \limits_{\beta \in
\mathfrak{B}}s^i_\beta f_\beta.\nonumber
\end{eqnarray}
But \(\forall \alpha \in \mathfrak{A}\) we have
\begin{eqnarray}
\sum \limits_{i=0}^\infty s_\alpha^n=\sum \limits_{i=0}^\infty
\langle e_\alpha \rangle_i=\langle e_\alpha \rangle=0,\nonumber
\end{eqnarray}
because \(e_\alpha \in V_s\) and \(\langle\cdot\rangle\) is a
translation invariant stationary functional. Therefore \(\forall
a\in V\)
\begin{eqnarray}
\langle a\rangle=\sum
\limits_{i=0}^{\infty}\langle\cdot\rangle_i=\sum
\limits_{i=0}^{\infty}(\sum \limits_{\alpha \in
\mathfrak{A}}s^i_\alpha f_\alpha(a) + \sum \limits_{\beta \in
\mathfrak{B}}s^i_\beta f_\beta(a))\nonumber\\
 =\sum
\limits_{i=0}^{\infty}\sum \limits_{\alpha \in
\mathfrak{A}}s^i_\alpha f_\alpha(a) + \sum
\limits_{i=0}^{\infty}\sum
\limits_{\beta \in \mathfrak{B}}s^i_\beta f_\beta(a)\nonumber\\
=\sum \limits_{\alpha \in \mathfrak{A}}(\sum
\limits_{i=0}^{\infty}s^i_\alpha) f_\alpha(a) + \sum
\limits_{i=0}^{\infty}\sum
\limits_{\beta \in \mathfrak{B}}s^i_\beta f_\beta(a)\nonumber\\
=\sum \limits_{i=0}^{\infty}(\sum \limits_{\beta \in
\mathfrak{B}}s^i_\beta f_\beta)(a)\nonumber
\end{eqnarray}
in Euler sense. Finally
\begin{eqnarray}
\langle a\rangle=\sum \limits_{i=0}^{\infty}\langle a\rangle'_i,\nonumber
\end{eqnarray}
where we put \(\langle\cdot\rangle'_i =\sum \limits_{\beta \in
\mathfrak{B}}s^i_\beta f_\beta \in V'_G\) and our statement is
proved.

Let 1 be some enough large but finite subsystem of our system. Let
2 be a subsystem obtained from 1 by translation on the vector
\(\vec{l}\) of sufficiently large length parallel to the
\(x\)-axis. Let 12 be a union of the subsystems 1 and 2. Let
\(U_1\), \(U_2\) and \(U_{12}\) be density matrices for the
subsystems 1, 2 and 12 respectively (which correspond to
\(\langle\cdot\rangle\)).

Let \(\{\varphi_{n,N}^1,n,N=1,2,...\}\) be a basis of eigenvectors
of Hamilton operator for subsystem 1.
Let\(\{\varphi_{m,M}^2,m,M=1,2,..\}\) be a basis of eigenvectors
of Hamilton operator for subsystem 2. \(N,M\) are the number of
particles in systems \(1,2\) respectively. Then the basis of
eigenvectors of Hamiltonian for subsystem 12 is
\(\{\psi_{n,N,m,M}=\varphi_{n,N}^1\otimes\varphi_{m,M}^2,\;n,N,m,M=1,2,...\}\).
According to the assumption that the systems 1,2 are enough large
we find the following expression
\begin{eqnarray}
U_{1,2}=\sum c_\alpha \frac{e^{-\frac{H_{1,2}-\mu_\alpha
N_{1,2}}{T_\alpha}}}{Z_\alpha}\nonumber
\end{eqnarray}
for density matrix of subsystems 1,2 in obvious notations. We have
also the following expression for the density matrix of subsystem
12 (if \(l=+\infty\)).
\begin{eqnarray}
U_{12}=\sum c_\alpha \frac{e^{-\frac{H_{1}-\mu_\alpha
N_1}{T_\alpha}}}{Z_\alpha}\otimes\frac{e^{-\frac{H_{2}-\mu_\alpha
N_2}{T_\alpha}}}{Z_\alpha}.\nonumber
\end{eqnarray}
But if \(l=+\infty\) the weak cluster property implies that
\(U_{12}=U_1\otimes U_2\). This leads to the following relations:
\begin{eqnarray}
\sum \limits_{\alpha} c_\alpha \frac{e^{-\frac{E_{n,N}-\mu_\alpha
N}{T_\alpha}}}{Z_\alpha}\frac{e^{-\frac{E_{m,M}-\mu_\alpha
M}{T_\alpha}}}{Z_\alpha}\nonumber\\
= \sum \limits_{\alpha} c_\alpha\frac{e^{-\frac{E_{n,N}-\mu_\alpha
N}{T_\alpha}}}{Z_\alpha} \sum \limits_{\beta}
c_\beta\frac{e^{-\frac{E_{m,M}-\mu_\beta M}{T_\beta}}}{Z_\beta}.\nonumber
\end{eqnarray}
Here \(\{E_{n,N},\;n,N=1,2,...\}\) be a set of eigenvalues of
\(H_{1,2}\). But the set of sequences
\begin{eqnarray}
\{e^{-\frac{E_{n,N}-\mu_\alpha N}{T_\alpha}}\}\nonumber
\end{eqnarray}
is linear independent if for all two indices \(\alpha,\beta\),
such that \(\alpha\neq\beta\) \((T_\alpha,\mu_\alpha)\neq
(T_\beta,\mu_\beta)\). So for each \(\alpha\) we have
\begin{eqnarray}
c_\alpha\frac{e^{-\frac{E_{n,N}-\mu_\alpha
N}{T_\alpha}}}{Z_\alpha}=c_\alpha\sum \limits_\beta c_\beta
\frac{e^{-\frac{E_{n,N}-\mu_\beta N}{T_\beta}}}{Z_\beta}.\nonumber
\end{eqnarray}
According to the linear independence of
\(\{e^{-\frac{E_{n,N}-\mu_\alpha N}{T_\alpha}}\}\) we find that
\(\forall \alpha\)
\begin{eqnarray}
c_\alpha=c_\alpha^2.\nonumber
\end{eqnarray}
So the series representing \(c_\alpha\) are convergent in the
Euler sense and \(\forall \alpha \;c_\alpha=0,1\).

But we have \(\sum \limits_\alpha c_\alpha={\rm tr \mit} U=1\). So
for some \(\beta\) \(c_\beta=1\) and \(c_\alpha=0\) if
\(\alpha\neq\beta\). We see that \(\langle\cdot\rangle\) is a
canonical Gibbsian distribution
\begin{eqnarray}
\langle\cdot\rangle=\langle\cdot\rangle_{\beta,\mu}\nonumber
\end{eqnarray}
for some inverse temperature \(\beta\) and chemical potential
\(\mu\).

Let us calculate now \(\langle a(k) a^+(k')\rangle\). We have
constructed \(\langle\cdot\rangle\) by some Gauss state \(\rho_0\)
described by some test function \(n(k)\). It follows from our
construction of \(\langle\cdot\rangle\) that:
\begin{eqnarray}
\langle a(k) a^+(k')\rangle=n(k)\delta(k-k').\nonumber
\end{eqnarray}
But if momentum \(k\) is sufficiently large we can neglect by
potential energy and find
\begin{eqnarray}
\langle a(k) a^+(k')\rangle=\rm const \mit
e^{-\beta\frac{k^2}{2}}\delta(k-k').\nonumber
\end{eqnarray}
If we chose \(n(k)\) such that \(n(k)\) tends to zero as
\(k\rightarrow \infty\) slowly than each Gauss function we obtain
a contradiction. This contradiction proves non ergodic property of
our system.

Now let us discuss so called the Boltzmann ergodic hypothesis
(1871). Let \(\langle\cdot\rangle\) be a translation invariant
stationary functional on \(V\) such that \(\forall t \in
\mathbb{R}\) the functionals \(\langle e^{itH}(\cdot)
e^{-itH}\rangle\) are well defined. The Boltzmann hypothesis
states that for each such functional there exists an element
\(\langle\cdot\rangle' \in V'_G\) such that \(\forall a \in V\)
\begin{eqnarray}
\lim \limits_{T\rightarrow +\infty}\frac{1}{T} \int
\limits_{0}^{T}\langle e^{itH} a e^{-itH}\rangle dt=\langle
a\rangle'.
\end{eqnarray}
We see that according to \(V'_s\neq V'_G\) the Boltzmann
hypothesis does not hold.
\section{Examples, chain diagrams}

In this section we consider by direct calculation some class of divergent diagrams in Keldysh diagram technique.
At first let us introduce the basis notion of the Keldysh diagram technique.

Let us introduce the Green functions for the system
\begin{eqnarray}
\rho(T(\Psi^\pm_H(t_1, x_1),...,\Psi^\pm_H(t_n, x_n))).\nonumber
\end{eqnarray}
Symbol \(H\) near \(\Psi^\pm\) means here that \(\Psi^\pm_H\) are
Heizenberg operators.

We require in nonequilibrium diagram technique the following
representation for the Green functions
\begin{eqnarray}
\rho(T(\Psi^\pm_H(t_1,x_1),...,\Psi^\pm_H(t_n,x_n)))=\nonumber\\
\rho_0(S^{-1}T(\Psi^\pm_0(t_1,x_1),...,\Psi^\pm_0(t_n,x_n)S)).\nonumber
\end{eqnarray}
The symbol \(0\) near \(\Psi^{\pm}\) means here that
\(\Psi^{\pm}_0\) are operators in the Dirac representation
(representation of interaction). The \(S\)-matrix has the form
\begin{eqnarray}
S=T\rm exp \mit{(-i\int \limits_{-\infty}^{+\infty}V(t)dt)},\;\nonumber
\end{eqnarray}
and
\begin{eqnarray}
S^{-1}=\tilde{T}\rm exp \mit{(i\int
\limits_{-\infty}^{+\infty}V(t)dt)}.\nonumber
\end{eqnarray}
\(\tilde{T}\) is a symbol of the antichronological ordering here.
\(\rho_0\) is some Gauss state defined by density function
\(n(k)\) as usual.

Let us recall the basic elements of nonequilibrium diagram
technique. The vertices coming from \(T\)-exponent are marked by
symbol \(-\). The vertices coming from \(\tilde{T}\)-exponent are
marked by symbol \(+\). There exist four types of propagators
\begin{eqnarray}
G_0^{+-}(t_1-t_2,x_1-x_2)=\rho_0(\Psi(t_1,x_1)\Psi^{+}(t_2,x_2)),\nonumber\\
G_0^{-+}(t_1-t_2,x_1-x_2)=\rho_0(\Psi^+(t_2,x_2)\Psi(t_1,x_1)),\nonumber\\
G_0^{--}(t_1-t_2,x_1-x_2)=\rho_0(T(\Psi(t_1,x_1)\Psi^+(t_2,x_2))),\nonumber\\
G_0^{++}(t_1-t_2,x_1-x_2)=\rho_0(\tilde{T}(\Psi(t_1,x_1)\Psi^+(t_2,x_2))).    \nonumber
\end{eqnarray}
Let us write the table of propagators
\begin{eqnarray}
G_0^{+-}(t,x)=\int \frac{d^4k}{(2\pi)^4}(2\pi)\delta(\omega-\omega(k))(1+n(k))e^{-i(\omega t-kx)},\nonumber\\
G_0^{-+}(t,x)=\int \frac{d^4k}{(2\pi)^4}(2\pi)\delta(\omega-\omega(k))n(k)e^{-i(\omega t-kx)},\nonumber\\
G_0^{--}(t,x)=i\int
\frac{d^4k}{(2\pi)^4}\{\frac{1+n(k)}{\omega-\omega(k)+i0}
-\frac{n(k)}{\omega-\omega(k)-i0}\}e^{-i(\omega t-kx)},\nonumber\\
G_0^{++}(t,x)=i\int
\frac{d^4k}{(2\pi)^4}\{\frac{n(k)}{\omega-\omega(k)+i0}
-\frac{1+n(k)}{\omega-\omega(k)-i0}\}e^{-i(\omega t-kx)}.\nonumber
\end{eqnarray}
\subsection{Divergences}
A typical example of divergent diagram is pictured at fig1.

\begin{picture}(300,100)
\put(10,80){fig. 1} \put(260,50){\vector(-1,0){45}}
\put(200,50){\oval(30,20)} \put(185,50){\vector(-1,0){15}}
\put(130,50){\vector(-1,0){15}} \put(85,50){\vector(-1,0){45}}
\put(100,50){\oval(30,20)}
\multiput(162,50)(-11,0){3}{\circle*{4}}
\put(260,45){\line(-1,0){6}} \put(218,41){+}
\put(224,39){\line(-1,0){6}} \put(176,41){+}
\put(176,39){\line(1,0){6}} \put(118,41){+}
\put(124,39){\line(-1,0){6}}
 \put(76,41){+}
\put(76,39){\line(1,0){6}} \put(40,45){\line(1,0){6}}
\end{picture}

The ovals represent the sum of one-particle irreducible diagrams.
These diagrams are called chain diagrams. Let us suppose that all
divergences of self-energy parts (ovals) are subtracted. The
divergences arise from the fact that singular supports of
propagators coincide. At first we consider diagrams with one
self-energy insertion (one-chain diagram). These diagrams are
pictured at fig. 2.

\begin{picture}(200,100)
\put(10,80){fig. 2}
\put(160,50){\vector(-1,0){45}}
\put(100,50){\oval(30,20)}
\put(85,50){\vector(-1,0){45}}
\put(160,45){\line(-1,0){6}}
\put(118,41){+}
\put(124,39){\line(-1,0){6}}
\put(76,41){+}
\put(76,39){\line(1,0){6}}
\put(40,45){\line(1,0){6}}
\end{picture}

 These diagrams are analogous to one-loop diagrams
in quantum field theory.

The aim of this section is to prove that the Green functions can be made finite by the following
renormalization of the asymptotical state:

\begin{eqnarray}
\rho_0(\cdot)\rightarrow \frac{1}{Z}\rho_0(e^{-\int \limits_{-\infty}^{+\infty} h(t)dt}(\cdot)),\nonumber
\end{eqnarray}
where
\begin{eqnarray}
h=\int h(k)a^+(k)a(k)d^3k,\nonumber
\end{eqnarray}
\(h(k)\) is a real-valued function and
\begin{eqnarray}
Z=\rho_0(e^{-\int \limits_{-\infty}^{+\infty} h(t)dt}).\nonumber
\end{eqnarray}

\subsection{Proof of the existence of divergences in the theory}

Suppose that there are no divergences in the Keldysh diagram technique
if \(n(k)\neq \frac{1}{e^{\alpha\frac{k^2}{2}+\beta}+1}\) for any positive \(\alpha,\;\beta\). Therefore
the Green function
\begin{eqnarray}
\rho( S^{-1}(S\Psi^+_0(t_1,x_1)\Psi_0(t_2,x_2)))\nonumber
\end{eqnarray}
is translation invariant. So the density matrix
\begin{eqnarray}
\rho_t(x_1,x_2):=\rho( S^{-1}(S\Psi^+_0(t,x_1)\Psi_0(t,x_2)))\nonumber
\end{eqnarray}
is an integral of motion. Let
\begin{eqnarray}
\rho_t(k)=\int d^3x \rho_t(0,x)e^{ikx}.\nonumber
\end{eqnarray}
In zero order of perturbation theory \(\rho(k)=n(k)\). But if
there are no divergences in Keldysh diagram technique it is
possible (see \cite{14}) to derive the following kinetic equation
for \(\rho(k)\)
\begin{eqnarray}
\frac{\partial \rho_t(k)}{\partial t}\nonumber\\
=\int w(p,p_1|p_2,p_3)\{(1+\rho(p))(1+\rho(p_1))\rho(p_2)\rho(p_3)\nonumber\\
-\rho(p)\rho(p_1)(1+\rho(p_2))(1+\rho(p_3))\}.\nonumber
\end{eqnarray}
The right hand side of this equation is equal to zero only if
\begin{eqnarray}
\rho(k)= \frac{1}{e^{\alpha\frac{k^2}{2}+\beta}+1}\nonumber
\end{eqnarray}
for some \(\alpha,\beta\) \((\alpha>0,\;\beta>0)\).
But \(n(k)=\rho(k)\) in zero order of perturbation theory, so
\(n(k)\) has a Bose-Einstein form. This contradiction proves our statement.
\subsection{Regularization} Let us now introduce regularization. Note that
\begin{eqnarray}
\frac{1}{x+i\varepsilon}=\frac{x}{x^2+\varepsilon}-\pi \frac{i}{\pi} \frac{\varepsilon}{x^2+\varepsilon^2}.\nonumber
\end{eqnarray}
Therefore we use the following regularization
\begin{eqnarray}
\delta(\omega-\omega(k))\rightarrow \frac{1}{\pi} \frac{\varepsilon}{(\omega-\omega(k))^2+\varepsilon^2}=:
\delta_\varepsilon(\omega-\omega(k)),
\nonumber\\
\mathcal{P}(\frac{1}{\omega-\omega(k)})\rightarrow \frac{{\omega-\omega(k)}}{(\omega-\omega(k))^2+\varepsilon^2}=:
\mathcal{P}_\varepsilon(\frac{1}{\omega-\omega(k)}).\nonumber
\end{eqnarray}
\subsection{Some simple relation on the Green functions}

\textbf{Lemma 1.} \textit{The following equalities hold}
\begin{eqnarray}
G^{--}(t_1-t_2,x_1-x_2)^\star=G^{++}(t_2-t_1,x_2-x_1),\label{I}\\
G^{+-}(t_1-t_2,x_1-x_2)^\star=G^{+-}(t_2-t_1,x_2-x_1).\label{II}
\end{eqnarray}

\textbf{Proof.} We have
\begin{eqnarray}
G^{--}(t_1-t_2,x_1-x_2)^\star=\rho_0(T(\Psi_H(t_1,x_1)\Psi^+_H(t_2,x_2)))^\star\nonumber\\
=\rho_0(\tilde{T}(\Psi^+_H(t_1,x_1)\Psi_H(t_2,x_2)))=G^{++}(t_2-t_1,x_2-x_1).\nonumber
\end{eqnarray}
So the equality \ref{I} is proved. We have
\begin{eqnarray}
G^{+-}(t_1-t_2,x_1-x_2)^\star=\rho_0(\Psi_H(t_1,x_1)\Psi^+_H(t_2,x_2))^\star\nonumber\\
=\rho_0(\Psi_H(t_2,x_2)\Psi^+_H(t_1,x_1))=G^{+-}(t_2-t_1,x_2-x_1).\nonumber
\end{eqnarray}
So the equality \ref{II} is proved.

The Lemma is proved.

It is easy to prove the following

\textbf{Lemma 2.} \textit{The following equality holds}
\begin{eqnarray}
G^{+-}(t,x)=\theta(t)G^{--}(t,x)+
\theta(-t)G^{++}(t,x),\nonumber\\
G^{-+}(t,x)=\theta(t)G^{++}(t,x)+
\theta(-t)G^{--}(t,x).\nonumber
\end{eqnarray}
Let us introduce the following matrix
\begin{eqnarray}
G=\left \|\begin{array}{cc}
 G^{++}&G^{+-}\\
 G^{-+}&G^{--}\\
\end{array} \right\|.\nonumber
\end{eqnarray}
Let us introduce the similar matrix for the self-energy operator
\begin{eqnarray}
\Sigma=\left \|\begin{array}{cc}
 \Sigma^{++}&\Sigma^{+-}\\
 \Sigma^{-+}&\Sigma^{--}\\
\end{array} \right\|.\nonumber
\end{eqnarray}
Dyson equations in Fourier representation have the form
\begin{eqnarray}
G=G_0+G_0\Sigma G.\nonumber
\end{eqnarray}
We have from these equations that
\begin{eqnarray}
\Sigma=G_0^{-1}-C^{-1},\nonumber
\end{eqnarray}
or in the matrix form
\begin{eqnarray}
\Sigma=\frac{1}{\rm det \mit G_0}\left \|\begin{array}{cc}
 G_0^{--}&-G_0^{+-}\\
 -G_0^{-+}&G_0^{++}\\
\end{array} \right\|-\frac{1}{\rm det \mit G}\left \|\begin{array}{cc}
 G^{--}&-G^{+-}\\
 -G^{-+}&G^{++}\\
\end{array} \right\|. \label{M}
\end{eqnarray}
It follows from Lemma 1 that
\begin{eqnarray}
G^{++}(\omega, p)=G^{--}(\omega, p)^\star \nonumber\\
G^{+-}(\omega, p)=G^{+-}(\omega, p)^\star, \nonumber\\
G^{-+}(\omega, p)=G^{-+}(\omega, p)^\star.\nonumber
\end{eqnarray}
Therefore \(\rm det \mit G_0,\; \rm det \mit G\) are real and we
have the following lemma.

\textbf{Lemma 3.}
\begin{eqnarray}
\Sigma^{--}(t_1-t_2,x_1-x_2)^\star=\Sigma^{++}(t_2-t_1,x_2-x_1),
\Sigma^{+-}(t_1-t_2,x_1-x_2)^\star=\Sigma^{+-}(t_2-t_1,x_2-x_1).
\end{eqnarray}

The following Lemma holds.

\textbf{Lemma 4.}
\begin{eqnarray}
\Sigma^{++}(\omega,p)+\Sigma^{--}(\omega,p)=-\Sigma^{-+}(\omega,p)-\Sigma^{+-}(\omega,p).\nonumber
\end{eqnarray}

\textbf{Proof.} The statement of lemma follows from the Dyson
equation (\ref{M}) and the following two obvious equalities:
\begin{eqnarray}
G^{++}(\omega,p)+G^{--}(\omega,p)=G^{-+}(\omega,p)+G^{+-}(\omega,p),\nonumber\\
G^{++}_0(\omega,p)+G^{--}_0(\omega,p)=G^{-+}_0(\omega,p)+G^{+-}_0(\omega,p).\nonumber
\end{eqnarray}

\subsection{Calculation of the propagators in one-chain approximation}
\textbf{Lemma 5.} \textit{The following limit equality holds (in the sense of distributions)}:
\begin{eqnarray}
\lim_{\varepsilon\rightarrow 0} (\delta_\varepsilon^2(x)-\frac{1}{2\pi\varepsilon}\delta_\varepsilon(x))=0,\nonumber \\
\lim_{\varepsilon\rightarrow 0}
(\frac{1}{\varepsilon}\delta_\varepsilon(x)-\frac{1}{\varepsilon}\delta(x))\rm
=reg \mit,
\nonumber\\
\lim_{\varepsilon\rightarrow 0}
\{\frac{1}{\pi}\frac{1}{x^2+\varepsilon^2}-\frac{1}{\varepsilon}\delta(x)\}=\rm
reg \mit,
\nonumber \\
\lim_{\varepsilon\rightarrow
0}\delta_\varepsilon(x)\mathcal{P}_\varepsilon(\frac{1}{x})\rm =reg
\mit,
\nonumber \\
x \in \mathbb{R}.\nonumber
\end{eqnarray}
Here reg  means some correct distribution.\nonumber\\
\textbf{Proof.} Let \(f(x)\) be some test function with compact support. We have

\begin{eqnarray}
\int \delta_\varepsilon^2(x)f(x)=\frac{1}{\pi^2} \int \frac{1}{(x^2+\varepsilon^2)^2}f(x)dx\nonumber\\
=\frac{1}{\pi^2}\int \frac{\varepsilon^2}{(x^2+\varepsilon^2)^2}
\{f(0)+xf'(0)+x^2\psi(x)\}dx\nonumber
\end{eqnarray}
for some smooth bounded function \(\psi(x)\). We have
\begin{eqnarray}
\int \delta_\varepsilon^2(x)f(x)=\frac{1}{\varepsilon \pi^2}
\int \frac{1}{(x^2+1)^2}\{f(0)+\varepsilon x f'(0)+\varepsilon^2x^2\psi(\varepsilon x)\}\nonumber\\
=\frac{1}{\pi^2} \{\frac{1}{\varepsilon} \int
\frac{1}{(x^2+1)^2}dx\}f(0)+O(\varepsilon).\nonumber
\end{eqnarray}
But
\begin{eqnarray}
\int \frac{1}{(x^2+1)^2}dx=\frac{\pi}{2}.\nonumber
\end{eqnarray}
So
\begin{eqnarray}
\int \delta_\varepsilon^2(x) f(x)=\frac{1}{2\pi \varepsilon} f(0)+O(\varepsilon).\nonumber
\end{eqnarray}
So the first equality is proved. One can prove other three equalities in the same way.

Therefore we see from the Lemmas 1,2, that we can consider only
the function \(G^{--}(t,x)\). But the function \(G^{--}(t,x)\) can
be represented as a sum of chain diagrams. At first let us
consider the diagrams with one self-energy insertion (one-chain
diagram). We have \(G_\varepsilon^{--}=\sum
\limits_{i,j=\pm}H^{ij}_\varepsilon\), where the diagrams
 for \(H^{ij}_\varepsilon\) are presented at the fig. 2.
  We have the following representation for the divergent parts of these diagrams.
\begin{eqnarray}
(H^{--}_\varepsilon)_{div}(\omega,p)+(H^{++}_{\varepsilon})_{div}(\omega,p)\nonumber\\
=2\pi \Sigma^{--}(\omega,p)n(p)(1+n(p))\frac{1}{\varepsilon}\delta(\omega-\omega(p))\nonumber\\
+2\pi
\Sigma^{++}(\omega,p)n(p)(1+n(p))\frac{1}{\varepsilon}\delta(\omega-\omega(p)).\nonumber
\end{eqnarray}
We see that the divergent part of these two diagrams is real
(because \(\Sigma^{--}=(\Sigma^{++})^*\)).

Let us
consider the singular part of other two diagrams presented at fig
3.

\begin{picture}(400,100)
\put(10,80){fig. 3}
\put(340,50){\vector(-1,0){45}}
\put(280,50){\oval(30,20)}
\put(265,50){\vector(-1,0){45}}
\put(340,45){\line(-1,0){6}}

\put(304,45){\line(-1,0){6}}
\put(256,41){+}

\put(220,45){\line(1,0){6}}

\put(160,50){\vector(-1,0){45}}
\put(100,50){\oval(30,20)}
\put(85,50){\vector(-1,0){45}}
\put(160,45){\line(-1,0){6}}
\put(118,41){+}

\put(76,45){\line(1,0){6}}
\put(40,45){\line(1,0){6}}
\end{picture}

We have
\begin{eqnarray}
(H_{\varepsilon}^{-+})_{div}(\omega,p)+(H_{\varepsilon}^{+-})_{div}(\omega,p)\nonumber\\
=\pi(2\pi)(1+2n(p))(1+n(p))\delta_\varepsilon^2(\omega-\omega(p))\Sigma^{-+}(\omega,p)\nonumber\\
+\pi(2\pi)(1+2n(p))n(p)\delta_\varepsilon^2(\omega-\omega(p))\Sigma^{+-}(\omega,p)\nonumber\\
=\pi(1+2n(p))(1+n(p))\frac{1}{\varepsilon}\delta(\omega-\omega(p))\Sigma^{-+}(\omega,p)\nonumber\\
+\pi(1+2n(p))n(p)\frac{1}{\varepsilon}\delta(\omega-\omega(p))\Sigma^{+-}(\omega,p)+O(\varepsilon).\nonumber
\end{eqnarray}
We see that \((H^{--}_\varepsilon)_{div}(\omega,p)+(H^{++}_\varepsilon)_{div}(\omega,p)\),
\((H^{-+}_\varepsilon)_{div}(\omega,p)\)+\((H^{+-}_\varepsilon)_{div}(\omega,p)\) are real.

We will use the dotted line for lines which connect creation-annihilation operators with operators
arising from the vertex: \(\int h(k) a^+(k)a(k)d^3k\) (see fig. 4).

\begin{picture}(200,100)
\put(10,80){fig. 4}
\put(120,50){\vector(-1,0){5}}
\multiput(160,50)(-10,0){5}{\circle*{2}}
\put(100,50){\oval(30,20)}
\put(45,50){\vector(-1,0){5}}
\multiput(85,50)(-10,0){5}{\circle*{2}}
\put(160,45){\line(-1,0){6}}
\put(118,41){+}
\put(124,39){\line(-1,0){6}}
\put(76,41){+}
\put(76,39){\line(1,0){6}}
\put(40,45){\line(1,0){6}}
\end{picture}

So the divergences in \(G^{--}_\varepsilon=\sum \limits_{i,j=\pm} H^{ij}_\varepsilon\)
can be subtracted  by the following counterterm:
\begin{eqnarray}
h(p)=\Sigma^{++}(\omega,p)+\Sigma^{--}(\omega,p)+\frac{(1+2n(p))}{2n(p)(1+n(p))} \nonumber\\
\times\{(1+n(p))\Sigma^{-+}(\omega,p)+n(p)\Sigma^{+-}(\omega,p)\}.\nonumber
\end{eqnarray}
By using Lemma 4 we have:
\begin{eqnarray}
h(p)=\frac{1+2n(p)}{2(1+n(p))n(p)}\times \nonumber\\
\{(1+n(p))\Sigma^{-+}(\omega,p)-n(p)\Sigma^{+-}(\omega,p)\}.\nonumber
\end{eqnarray}
The left hand side of this equation can be rewritten as follows
(in approximation used in \cite{14})
\begin{eqnarray}
h(p)=\frac{1+2n(p)}{2n(p)(1+n(p))} St(p),\nonumber
\end{eqnarray}
where \(St(p)\) is a scattering integral.
So \(h(p)\neq 0\) for non-equilibrium matter.

Analogously one can consider two-chain diagrams presented at fig. 5 by direct calculation and
prove that the divergences can be subtracted by the counterterms of the asymptotical state.

\begin{picture}(300,100)
\put(10,80){fig. 5}
\put(235,50){\vector(-1,0){45}}
\put(175,50){\oval(30,20)}
\put(85,50){\vector(-1,0){45}}
\put(100,50){\oval(30,20)}

\put(235,45){\line(-1,0){6}}
\put(193,41){+}
\put(199,39){\line(-1,0){6}}
\put(151,41){+}
\put(151,39){\line(1,0){6}}
\put(160,50){\vector(-1,0){45}}
\put(118,41){+}
\put(124,39){\line(-1,0){6}}
\put(76,41){+}
\put(76,39){\line(1,0){6}}
\put(40,45){\line(1,0){6}}
\end{picture}

Note that there arises the phenomenum of overlapping divergences in this example.

\section{Notes on Bogoliubov derivation of Boltzmann equations}
 In this section we study the problem of boundary conditions in Bogoliubov derivation of
 kinetic equations \cite{6}.
Let us consider \(N\) particles in \(\mathbb{R}^3\). Let \({q}_i\)
be a coordinates of particle number \(i\), and \({p}_i\) be a
momentum of particle number \(i\), \(i=1,...,N\). Suppose that
particles interact by means of the pair potential
\(V({q}_i-{q}_j)\). We suppose that \(V\) belongs to the Schwartz
space. Let \(x_i=({p}_i,{q}_i)\) be a point in the phase space
\(\Gamma\). Let \(f(x_1,...,x_n)\) be a distribution function of
\(N\) particles. If we want to point out that \(f(x_1,...,x_N)\)
depends on \(t\) we will write \(f(x_1,...,x_N|t)\). Let
\begin{eqnarray}
f_1(x_1)=\int dx_2...dx_N f(x_1,...,x_n),\nonumber
\end{eqnarray}
and
\begin{eqnarray}
f_2(x_1,x_2)=\int dx_3,...,dx_N f(x_1,...,x_N)\nonumber
\end{eqnarray}
be marginal distribution functions. Put by definition
\begin{eqnarray}
\rho_1(x_1)=Nf_1(x_1),\nonumber\\
\rho_2(x_1,x_2)=N^2f_2(x_1,x_2).\nonumber
\end{eqnarray}
If \(A\) is a function on the phase space \(\Gamma\),
\(\Gamma=\mathbb{R}^{6N}\) and
\begin{eqnarray}
A=\sum \limits_{i=1}^{N}\mathcal{A}(x_i)\nonumber
\end{eqnarray}
then
\begin{eqnarray}
\langle A\rangle=\int f(x_1,...,x_n) A(x_1,...,x_N)=\nonumber\\
N\int dx_1\mathcal{A}(x_1)f(x_1)=\nonumber\\
=\int dx \mathcal{A}(x)\rho_1(x).\nonumber
\end{eqnarray}
Now if \(A\) is a function on the phase space
\begin{eqnarray}
A=\sum \limits_{i\neq j}\mathcal{A}(x_i,x_j)\nonumber
\end{eqnarray}
in the limit of large \(N\) we find
\begin{eqnarray}
\langle A\rangle=\int dx_1 dx_2\rho_2(x_1,x_2)\mathcal{A}(x_1,x_2).\nonumber
\end{eqnarray}
Let us introduce also three-particle distribution function:
\begin{eqnarray}
f_3(x_1,x_2,x_3)=\int dx_4...dx_N f(x_1,...,x_N).\nonumber
\end{eqnarray}
Let us derive the equation for \(f(x)\). At first let us write
equation of motion for \(f(x_1,...,x_N)\). We have
\begin{eqnarray}
\frac{\partial}{\partial t}f(x_1,...,x_N\mid t)+\sum
\limits_{i=1}^{n}\frac{{p}_i}{m}{\nabla}_i
f(x_1,...,x_n|t)\nonumber\\
-\sum \limits_{i\neq j} \frac{{\partial} V(q_i-q_j)}{\partial
{q}_i}\frac{\partial f(x_1,...,x_n|t)}{\partial {p}_i}=0.\nonumber
\end{eqnarray}
This equation is only an infinitesimal form of the Liouville
theorem. Let us multiply this equation by \(N\) and integrate over
\(dx_2,...,dx_N\). Suppose that \(f(x_1,...,x_N)\) is a function
of rapid decay of momenta. This assumption allow integrate over
\(p_i\) by parts. We find:
\begin{eqnarray}
\frac{\partial}{\partial
t}\rho_1(x_1|t)+\frac{{p}}{m}{\nabla}\rho_1(x_1,t)\nonumber\\
+\int dx_2 \frac{{p_2}}{m}\frac{\partial}{\partial
{q}_2}\rho_2(x_1,x_2|t)\nonumber\\
=\int dx_2\frac{\partial V(q_1-q_2)}{\partial {q}_1}\frac{\partial
\rho_2(x_1,x_2|t)}{\partial p_1}. \label{0}
\end{eqnarray}
Note that we have kept here boundary term. Let us now talk about
derivation of kinetic equation. According to the standard prescription
we put \(\rho_3(x_1,x_2,x_3)=0\) in equation for
\(\rho_2(x_1,x_2)\). We find the following equation for
\(\rho_2\):
\begin{eqnarray}
\frac{d}{dt}\rho_2(x_1(t),x_2(t)|t)=0,\label{1}
\end{eqnarray}
where \((x_1(t),x_2(t))\) is a solution of corresponding two-body
problem.

\textbf{Condition of correlation breaking.} We consider only translation-invariant matter in purpose of simplicity. Usual
correlation-breaking condition has the form
\begin{eqnarray}
\rho_2(x_1,x_2|0)=h({p}'_1(x_1,x_2))h({p}'_2(x_1,x_2)).\nonumber
\end{eqnarray}
Here \(h\) is a function on momenta-space of one particle. We
consider only translation-invariant gas, so \(h\) depends only of
momentum.

\({p'}_1(x_1,x_2)\) and \({p'}_2(x_1,x_2)\) are momenta of
particles 1 and 2 at \(t=-\infty\) if at \(t=0\) their coordinates
and momenta was \(x_1\) and \(x_2\) respectively.

\textbf{Proposition.}
\begin{eqnarray}
\frac{\partial}{\partial t}\rho_2(x_1,x_2)=0.\label{2}
\end{eqnarray}
 Indeed, according to (\ref{1})
 \begin{eqnarray}
 \rho_2(x_1,x_2|t)=\rho_2(x_1^0,x_2^0|0),\nonumber
 \end{eqnarray}
 where \(x_1^0\) and \(x_2^0\) are phase coordinates of particles
 1 and 2 respectively at a moment \(t=0\). Therefore
 \begin{eqnarray}
\rho_2(x_1,x_2|t)=h({p}'_1(x^0_1,x^0_2))h({p}'_2(x^0_1,x^0_2)).\nonumber
\end{eqnarray}
But the points \(x_1^0\) and \(x_2^0\) come to the points \(x_1\)
and \(x_2\) after the time t. So \(({p}'_1(x^0_1,x^0_2)
,{p}'_2(x^0_1,x^0_2))=({p}'_1(x_1,x_2) ,{p}'_2(x_1,x_2))\), and
\begin{eqnarray}
\rho_2(x_1,x_2|t)=h({p}'_1(x^0_1,x^0_2))h({p}'_2(x^0_1,x^0_2))=\nonumber\\
h({p}'_1(x_1,x_2))h({p}'_2(x_1,x_2))=\rho_2(x_1,x_2|0).\nonumber
\end{eqnarray}
In result
\begin{eqnarray}
\rho_2(x_1,x_2|t)=\rho_2(x_1,x_2|0).\nonumber
\end{eqnarray}
The proposition is proved.

It follows from equations (\ref{1}) and (\ref{2}) that:
\begin{eqnarray}
(\frac{{p}_1}{m}{\nabla}_1+\frac{{p}_2}{m}{\nabla}_2)f_2(x_1,x_2|t)\nonumber\\
=(\frac{\partial V(q_1-q_2)}{\partial
{q}_1}\frac{\partial}{\partial {p}_1}+\frac{\partial
V(q_1-q_2)}{\partial {q}_2}\frac{\partial}{\partial
{p}_2})f_2(x_1,x_2|t). \label{3}
\end{eqnarray}
The function \(h(p)\) can be found from the following equation
\begin{eqnarray}
\rho_1(x)=\lim_{N\rightarrow\infty} \frac{1}{N}\int
\rho_2(x_1,x_2) dx_2.\nonumber
\end{eqnarray}
But in zero order of gas parameter the particles are free and
\begin{eqnarray}
\rho_1(x)=h(x).\nonumber
\end{eqnarray}
Formula (\ref{3}) is usually used for transformation of r.h.s. of
equation (\ref{0}) to the scattering integral. From other hand the
equation
\begin{eqnarray}
\frac{\partial}{\partial t}\rho_2(x_1,x_2)=0\nonumber
\end{eqnarray}
shows that there is no irreversible evolution in the system. From
other point of view we will show that the last term in the left
hand side of (\ref{0}) is equal to the scattering integral.

For simplicity we will show the case \(v_1=0\), \(v=\frac{p}{m}\).
The general case can be reduced to this case by means of Galilei
transformation. So let us consider the integral
\begin{eqnarray}
I=\lim_{R\rightarrow\infty} \int d^3{p}_2 \int
\limits_{V_R}d{q}_2\frac{{p}_2}{m}\frac{\partial}{\partial
{q}_2}\rho_2(0,0,{p}_2,{q}_2),\nonumber
\end{eqnarray}
where \(V_R\) is a ball of radius \(R\) with the center at zero.
Let us integrate over \(dq_2\) by using Gauss theorem. We find
\begin{eqnarray}
I=\lim_{R\rightarrow\infty}\int d^3p_2\int \limits_{S_R} dS
\frac{p_2}{m}\cos \psi \rho_2(0,0,{p}_2,{q}_2).\nonumber
\end{eqnarray}
Here \(S_R\) is a boundary of \(V_R\) and \(\psi\) is an angle
between two rays: first of them is parallel to \({p}_2\), second
starts from zero and passes throw \(q_2\). We have
\begin{eqnarray}
I=\lim_{R\rightarrow\infty} \int d^3p_2 \int \limits_{S_R}
dS\frac{p_2}{m}\cos\psi\times\nonumber\\
h({p}_1'(0,0))h({p}_2'({p}_2,{q}_2)). \nonumber
\end{eqnarray}
Let us suppose that the particles scatter only then they are not
too far from each other. Then
\begin{eqnarray}
h({p}_1'(0,0))h({p}_2'({p}_2,{q}_2))=\rho_1({p}_2)\rho_1(0)\nonumber
\end{eqnarray}
for all \({q}_2 \in S_R\setminus\mathcal{O}\), where
\(\mathcal{O}\) is a small neighborhood of the point
\({q}_0:=\frac{{p}_2}{|{p}_2|}R \in S_R\). Diameter of
\(\mathcal{O}\) is approximately equal to diameter of \(\rm supp
\mit V\). Therefore the integral \(I\) is not equal to zero and
equal to
\begin{eqnarray}
I=\int d^3p_2 \frac{p_2}{m} \int 2\pi b db \nonumber\\
\times\{\rho_1({p}_1'({p}_2,{q}_2({b})),(0,0))
\rho_1({p}_2'({p}_2,{q}_2({b})),(0,0))\nonumber\\
-\rho_1({p}_2)\rho_1(0)\}, \label{5}
\end{eqnarray}
where \({b}:={q}_2-{q}_0\). But the right hand side of (\ref{5})
is a usual scattering integral.

Therefore if we keep boundary terms in BBGKI-chain we obtain the
kinetic equations without scattering integral.
\section{Conclusion}
In the present paper we have developed the general theory of the
renormalization of nonequlibrium diagram technique. To study this
problem we have used some ideas of the theory of \(R\)-operation
developed by N.N. Bogoliubov and O.S. Parasiuk.

We illustrate our ideas by simple example of one- and two-chain
diagrams in Keldysh diagram technique.

We want to illustrate in this paper the following general thesis:
to prove that the system tends to the thermal equilibrium one
should take into account its behavior on its boundary. In the last
section we have shown that some boundary terms in BBGKI-chain
which are usually neglected in Bogoliubov derivation of kinetic
equation compensate scattering integral in kinetic equation.

Author is grateful to I.V. Volovich, O.G.
Smolyanov, Yu. E. Lozovik, A.V. Zayakin and I.L. Kurbakov for very
useful discussions.

\end{document}